# The De-democratization of AI: Deep Learning and the Compute Divide in Artificial Intelligence Research

**Nur Ahmed**[*]

**Muntasir Wahed**[ξ]

## Abstract:

Increasingly, modern Artificial Intelligence (AI) research has become more computationally intensive. However, a growing concern is that due to unequal access to computing power, only certain firms and elite universities have advantages in modern AI research. Using a novel dataset of 171,394 papers from 57 prestigious computer science conferences, we document that firms, in particular, large technology firms and elite universities have increased participation in major AI conferences since deep learning's unanticipated rise in 2012. The effect is concentrated among elite universities, which are ranked 1-50 in the QS World University Rankings. Further, we find two strategies through which firms increased their presence in AI research: first, they have increased firm-only publications; and second, firms are collaborating primarily with elite universities. Consequently, this increased presence of firms and elite universities in AI research has crowded out mid-tier (QS ranked 201-300) and lower-tier (QS ranked 301-500) universities. To provide causal evidence that deep learning's unanticipated rise resulted in this divergence, we leverage the generalized synthetic control method, a data-driven counterfactual estimator. Using machine learning based text analysis methods, we provide additional evidence that the divergence between these two groups—large firms and non-elite universities—is driven by access to computing power or compute, which we term as the "compute divide". This compute divide between large firms and non-elite universities increases concerns around bias and fairness within AI technology, and presents an obstacle towards "democratizing" AI. These results suggest that a lack of access to specialized equipment such as compute can de-democratize knowledge production.

[*] Ivey Business School, Western University.

[ξ] Virginia Tech.

[ψ] Correspondence: nahmed@ivey.ca; We are grateful to JP Vergne, Neil C. Thompson, Abhishek Nagaraj, Romel Mostafa, Subrina Shen, Leo Schmallenbach, Brandon Schaufele, Mayur P. Joshi, Andrew Sarta, Morgan Frank, and seminar participants at Industry Studies Association, Western Computational Social Science, Ivey Research Series. Farhana Zaman, Fajria Mannan, and Salsabil Tarannum have provided excellent research assistance. All errors are our own.

> "[ In AI…] Currently the power, the expertise, the data are all concentrated in the hands of a few companies" — **Yoshua Bengio**, 2018 Turing award recipient and professor, University of Montreal (Murgia, 2019)

## Introduction

Artificial Intelligence (AI) has been labeled a "general purpose technology" due to its pervasive role across many different industries (Cockburn, Henderson, & Stern, 2018), and has substantial implications for innovation and socio-economic development (Cockburn et al., 2018; Goolsbee, 2018; Korinek & Stiglitz, 2017). However, researchers argue that this technology is increasingly being deployed in highly consequential domains such as education, healthcare, and criminal law without taking into account its potential consequences (West, Whittaker, & Crawford, 2019). A large body of work raises significant concerns about biases and fairness in AI-enabled technologies and its underlying algorithms and datasets (Bolukbasi, Chang, Zou, Saligrama, & Kalai, 2016; Buolamwini & Gebru, 2018; Delmestri & Greenwood, 2016; Righetti, Madhavan, & Chatila, 2019). More specifically, AI has the potential to change the existing social order by replacing jobs or predicting hiring decisions. Thus, understanding who designs and shapes this technology is of paramount importance.

Due to the pervasive presence of AI, researchers and policymakers have advocated for inclusive AI, and there is a growing consensus to "democratize" AI to ensure that the benefits of this technology are not limited to a small group of people (Fei-Fei, 2018; Knight, 2018; Kratsios, 2019). Democratizing AI can be defined as "making it possible for everyone to create artificial intelligence systems" (Riedl, 2020). Democratizing AI is vital because when AI is biased and not fair, it has the potential to exacerbate existing social inequities. To mitigate biases and unfairness in AI, scholars have underscored the importance of diversity among AI researchers (Kuhlman, Jackson, & Chunara, 2020; West et al., 2019). A diverse representation among researchers increases the possibility that different domain expertise and diverse perspectives and experiences will contribute to mitigating biases in proposed models and datasets.

However, there is a growing concern that AI research is becoming less democratized and more concentrated due to increased industry presence and lack of access to resources (Frank, Wang, Cebrian, & Rahwan, 2019; Metz, 2017; Murgia, 2019; Riedl, 2020; The Economist, 2016b). Due to the computational nature of "modern AI" or "post-Moore" era AI research, where available computing power or compute is doubling at a much faster rate than before, the concern is that only a small group of actors will shape the future of AI. The modern AI's remarkable success across many different fields, such as object recognition, machine translation, text generation, and other areas (Shoham et al., 2019), has been primarily driven by "deep learning" (LeCun, Bengio, & Hinton, 2015). Deep learning is an improved version of neural networks, a decades-old algorithm that crudely mimics the human brain. The remarkable success of modern AI has been possible, in part, due to the availability of enormous computing power and large datasets (Gupta, Agrawal, Gopalakrishnan, & Narayanan, 2015; Thompson & Spanuth, 2018). However, anecdotal evidence of uneven access to compute has increased concerns within the AI community (Amodei & Hernandez, 2018; Riedl, 2020). For instance, the New York Times speculated: *"[…] pioneering artificial intelligence research will be a field of haves and have-nots. And the haves will be mainly a few big tech companies like Google, Microsoft, Amazon and Facebook, which each spend billions a year building out their data centers*" (Lohr, 2019). While recent evidence (Frank et al., 2019) also indicates increased concentration within AI research, until now, there has not been any systematic study on large organizations, in particular, large firms' dominance in modern AI. This study examines these concerns systematically to test whether a systemic divergence exists between different groups in AI research. Specifically, we ask *(i) Are we observing an increased concentration of AI research among a few actors since deep learning's rise? (ii) Who are the key contributors to "modern AI" research? (iii) What are the implications for organizations that have been previously active in AI research?*

To answer these questions, we construct a novel and extensive dataset of 171,394 papers from 57 leading computer science academic venues, both AI (e.g., Computer Vision, Machine Learning & Data mining, and Natural Language Processing) and non-AI conferences (e.g., Human-Computer Interaction, Software Engineering, Mobile Computing). These conferences are largely similar in terms of selectivity, prestige, and impact, hence shape the broader computer science research (Freyne, Coyle, Smyth, & Cunningham, 2010). To measure the causal effect of deep learning's unanticipated rise on organizational participation in AI research, we exploit the ImageNet contest's 2012 edition as a shock. The scientific community did not anticipate the widespread use and impact of Graphics Processor Units (GPUs) for deep learning research. However, the winners of the ImageNet 2012's contest demonstrated that GPU-trained deep learning models produce superior results, which led to the sudden popularity of this method. The unanticipated rise of deep learning provides an exogenous event that is correlated with increased compute due to increased GPU use, but not with research organizations' research behavior. However, GPU usage affects research behavior indirectly through the increased compute that allows effective training of AI algorithms, and, in turn, has produced superior research outputs. Exploiting this fact, we create a valid counterfactual by using a recently developed counterfactual estimator (Liu, Wang, & Xu, 2019)—generalized synthetic control method (Xu, 2017). This method creates reliable counterfactuals for the treated units by using non-treated or control units. In this case, using major non-AI conferences (which have not been affected much by the deep learning revolution), the methods create a "synthetic" counterfactual for major AI conferences (which have been significantly affected by the deep learning revolution) for the treatment periods.

Using the generalized synthetic control method and an extensive dataset, we present systematic evidence that firms, in particular, large technology firms, are increasingly contributing more to AI research relative to other computer science areas. Our estimates suggest that Fortune 500 Global technology firms are publishing 44 additional papers annually per AI conference than the counterfactual. This is a significant change given that these firms' average annual publication is only 23 papers per computer science conference. Similarly, elite universities (QS ranked 1-50) are publishing 40 additional papers per year per conference. In contrast, mid-tier universities (QS ranked 201-300) and lower tier universities (QS ranked 301-500) are publishing 14 and 5 fewer papers, respectively. Additionally, we document that Historically Black Colleges and Universities (HBCU), and Hispanic-serving institutions (Hispanic Association of College and Universities or HACU) are underrepresented in top AI venues.

Furthermore, using frequency-inverse document frequency or TF-IDF analysis, we provide evidence that the growing divergence in AI knowledge production between non-elite universities and large technology firms is attributable, in part, to the increasing divide in access to compute. We term this uneven distribution in access to computing power the "compute divide." Our text analysis suggests that large technology firms are publishing more in deep learning areas than both elite and non-elite universities.

We make two important contributions to the innovation literature. First, we contribute to the vibrant and emerging literature on the role of research materials and equipment in knowledge production (Ding, Levin, Stephan, & Winkler, 2010; Furman & Teodoridis, 2020; Murray, Aghion, Dewatripont, Kolev, & Stern, 2016). Using a simple model of scientific knowledge production, we predict and then find support for the hypothesis that the compute-intensive nature of modern AI has resulted in de-democratization (Stephan, 2012). To the best of our knowledge, this is the first study that finds evidence that an increased need for specialized equipment can result in "haves and have-nots" in a scientific field.

Second, we document an important trend in corporate science that contradicts recent innovation research. While recent evidence in innovation literature suggests that firms have reduced corporate research (Arora, Belenzon, & Patacconi, 2018; Larivière, Macaluso, Mongeon, Siler, & Sugimoto, 2018), our study provides evidence that with respect to AI, firms have increased corporate research significantly. This opens up new opportunities for researchers to examine the reasons behind sudden increased corporate presence in AI. Furthermore, we also document the role of specialized equipment (e.g., data, compute) in facilitating increased collaboration between firms and universities. The increased corporate presence in AI research is driven by firm-level increased AI publications, on the one hand, and collaboration between firms and elite universities on the other. The increased collaboration between firms and elite universities is potentially explained by their complementary resources, where firms have access to compute and proprietary datasets, and elite universities have trained scientists.

Our results have important implications for policymakers and innovation scholars. First, the crowding out of a large number of universities is concerning because in computer science, conferences are the primary venue for shaping the research direction (Freyne et al., 2010). Our results suggest that rather than the democratization of AI, we observe a concentration of knowledge production by a small group of actors or the de-democratization of AI due to the compute divide. In contrast, in other academic areas such as life sciences and economics, non-elite universities are catching up in research with elite universities (Agrawal & Goldfarb, 2008; Halffman & Leydesdorff, 2010; Kim, Morse, & Zingales, 2009). Our results find the first concrete evidence that governments may have to step up to reduce the compute divide by providing a "national research cloud" as recently advocated by computer scientists (Walsh, 2020).

Second, the combined effect of increased industry presence and the crowding out of non-elite universities may have long-term consequences for the future direction of AI research. Scholars argue that AI firms are less diverse than academia (West et al., 2019) and produce research that is less reproducible than academic research (Gundersen & Kjensmo, 2018). AI firms are often more interested in commercial research (Frank et al., 2019; Hooker, 2020; Murgia, 2019). In other words, the increased presence of firms may have negative consequences for long-term innovation. Furthermore, it is well-documented that elite universities are also racially less diverse, and represent mostly wealthy families relative to non-elite universities (Chetty, Friedman, Saez, Turner, & Yagan, 2019; Reardon, Baker, & Klasik, 2012). In contrast, scholars underscore the importance of having diversity among AI researchers to mitigate biases (Abebe, 2018; Kuhlman et al., 2020; West et al., 2019). Our findings suggest that modern AI is being increasingly shaped by a small number of organizations that are less diverse than the broader population. Taken together, our results emphasize the need for research on the antecedents and consequences of the profound shift in knowledge production in AI.

## Relevant Literature

### The Modern AI Research: The Increased Role of Compute

Computer science research depends on a combined effect of algorithms, hardware (or compute), and specialized software for that hardware. More specifically, throughout AI's entire history, compute has played an important role in determining what counts as a breakthrough (Sutton, 2019) and which direction the research follows (Hooker, 2020). In fact, researchers argue that available compute can play a role of "lottery" in deciding which research direction scientists pursue. In other words, rather than the algorithms or software, compute can play an outsize role in determining the direction of the research (Hooker, 2020).

Accordingly, computer scientists divide the entire history of AI research from the 1950s through 2019 into two distinct eras based on compute usage (Amodei & Hernandez, 2018; Shoham et al., 2019). The first era is from the 1950s until 2011, where compute-usage followed Moore's law. Simply put, Moore's law suggests that the available compute doubles every two years. In this era, research was driven by general purpose hardware (Hooker, 2020; Thompson & Spanuth, 2018). Every researcher used the same hardware because they expected that the available compute would continue to increase at a predictable rate. During this period, investing in specialized hardware was relatively riskier, given specialized hardware requires additional investment and resources (Feldman, 2019) and there is no certainty that the investment will pay off. Therefore, in this era, AI scientists focused on research with general purpose hardware (Hooker, 2020).

In contrast, the second era, or what scientists call the "modern era", starting in 2012, is characterized by compute-usage significantly outpacing the previous period (Amodei & Hernandez, 2018). During the modern era, compute-usage has been doubling every 3.4 months rather than every two years. This increase is primarily due to specialized hardware and usage of unconventional processing units such as GPUs to train models. While GPUs were popular before 2012 for video games or developing graphics for animation, the technology was repurposed in the late 2000s for AI research. In the 2010s large firms started to invest more in specialized hardware such as Google's Tensor Processing Unit (TPUs) and Facebook's Big Sur (Lee & Wang, 2018). Thus, since 2012, AI research has transitioned from general purpose hardware to specialized hardware.

This profound shift in hardware also changed the AI research landscape. Before 2012, every researcher relied primarily on general purpose hardware or CPUs. Accordingly, during that period, most researchers used the same software and hardware. Therefore, other things being equal, researchers were mostly competing on the superiority of their algorithms (or ideas). However, in modern AI, due to the availability of specialized hardware, researchers are not on equal footing. In particular, when compute plays an outsize role in determining which ideas are better, competing to produce research becomes more difficult for scientists who do not have access to specialized hardware. In sum, in modern AI, specialized hardware or increased compute provides a competitive advantage to certain research groups.

**The Role of Compute, Data, and Human Capital in Modern AI Research**

In deep learning driven modern AI research, trained scientists, data and compute play important roles in deciding who can participate in top-tier research venues. Large technology firms and elite universities have multiple advantages over non-elite universities in modern AI research. First, firms have greater resources to recruit talent from universities. Human capital is particularly important in modern AI research, where the dominant method is deep learning. Deep learning research is heavily reliant on people who have accumulated related expertise through formal training (e.g., Ph.D.) or years of applied work. This is because researchers still have a limited understanding of why deep learning works, which partly explains the lack of codification in this area (Chen, 2019; Martineau, 2019). Overall, the lack of understanding and codification makes the diffusion of deep learning research much harder, and thus, human capital plays a more significant role in producing new knowledge compared to other computer science subfields.

The gap between the limited supply and the increased demand for AI talent has produced a talent scarcity in the field (Metz, 2018). Due to the shortage of talent in AI, qualified scientists are compensated millions of dollars in annual salary (Metz 2018, Economist 2016). Most universities cannot afford to pay such steep remuneration to AI scientists. Consequently, recent evidence suggests that large firms with deeper pockets are recruiting faculty members away from universities (Gofman & Jin, 2019; The Economist 2016; Murgia, 2019). For instance, Gofman and Jin (2019) document that between 2004 and 2018, large firms have recruited 180 faculty members from North American universities. Furthermore, large technology firms have acquired dozens of startups over the last eight years, primarily to acquire AI talent (Bass & Brustein, 2020). In sum, firms have an advantage over universities in recruiting and retaining AI scientists.

Second, large firms have access to compute and unique proprietary large datasets, which are important ingredients for modern AI research (Gupta et al., 2015; Strubell, Ganesh, & McCallum, 2019). Specifically, compute played a major role in deep learning's superior performance (Thompson, Greenewald, Lee, & Manso, 2020). Later research also demonstrated that increased compute leads to better performance and is often complementary to algorithms (Amodei & Hernandez, 2018; Hestness et al., 2017; Shazeer et al., 2017). However, this increased compute requires significant investment in relevant technologies such as GPUs or compute cloud. For instance, DeepMind's AlphaGo Zero, which was self-trained and able to outperform the previous record-holder AlphaGo, was trained at a cost estimated at $35 million. For comparison, the world's largest robotics lab at Carnegie Mellon University has an annual budget of $90 million (Schackner, 2019). Furthermore, recent innovations in designing chips and cloud computing have made it possible for large firms to get ahead in the competition. For example, firms such as Amazon, Apple, Google, and Tesla have been designing specialized chips. In sum, AI research has moved into the post-Moore computing era, where general purpose chips do not improve with time; this situation benefits only a smaller group of organizations that can design specialized chips and write specific software for the hardware (Thompson & Spanuth, 2018).

Finally, firms have quality proprietary datasets that contribute to better training datasets which produce highly accurate deep learning models. Recent research suggests that large firms like Facebook, Google, and Amazon have an advantage in AI research due to their proprietary data (Shokri & Shmatikov, 2015; Traub, Quiané-Ruiz, Kaoudi, & Markl, 2019). However, both elite and mid-tier universities lack access to compute and large datasets.

In sum, due to the AI industry's significant resources, large firms have been able to recruit and retain top talent, representing an important resource for AI research. Further, firms have advantages in AI because they have access to large proprietary datasets and compute. Overall, the White House's 2019 AI report summarized the central problem as follows: *"[...] industry, with its sustained financial support and access to advanced computing facilities and datasets, exerts a strong pull on academic research and teaching talent"* (Kratsios, 2019: 37).

**Research Equipment and the De-Democratization of Knowledge Production**

After discussing how modern AI is particularly dependent on human capital, compute, and data, we rely on economics of science literature to build a simple model to predict the potential consequences of the rise of deep learning on organizational participation. Economics of science research suggests that knowledge production depends on a number of factors such as specialized equipment and materials. Following prior research (Ding et al., 2010; Stephan, 2012), we use a simple model of knowledge production to formalize how assets like datasets and compute play a role. We assume that scientific knowledge production depends on knowledge, skills, materials, equipment, and effort.

KP= f (knowledge, skills, materials, equipment, and effort) (1)

Where KP is a measure of output such as the number of publications at the organizational level and the five factors are inputs in this equation. Let's assume that the marginal product of each argument is positive. We assume that organizations do not have to create or supply these inputs directly but can rely on other organizations. For instance, a university can collaborate with a firm to utilize the firm's specialized equipment. The equation suggests that other things being equal, an increase or decrease in any input would increase and decrease the KP, respectively.

This model is aligned with prior research (Murray et al., 2016; Nagaraj, Shears, & de Vaan, 2020; Teodoridis, 2018) that acknowledges the importance of materials and specialized equipment in producing new knowledge. In particular, literature suggests that access to research tools and data can have a democratizing effect on knowledge production. For instance, Nagaraj et al. (2018) argue that access to data democratizes science by allowing a geographically diverse set of actors to explore a diverse set of topics. In the same vein, Teodoridis (2018) demonstrated that reducing the cost of a research technology can facilitate collaboration with experts outside of the focal field to produce new knowledge. However, until now, there is no empirical study on how the increased cost of doing research can "de-democratize" a scientific field.

We contend that the rise of deep learning increases the importance of compute and data drastically, which, in turn, heightens the barriers of entry by increasing the costs of knowledge production. In her seminal book "How Economics Shapes Science", Paula Stephan (2012:108) hypothesized that *"[...] the increased importance of equipment and the high costs of equipment can increase the disparity between the haves and have-nots.*" Organizations that have limited access to such equipment struggle to produce new knowledge, whereas organizations that have access to such equipment advance in scientific knowledge production. Thus, in our setting, due to the rise of deep learning, we should observe a disparity in modern AI research.

Moreover, the drastically increased need for specialized equipment and materials in modern AI (e.g., compute, dataset) is unlikely to affect universities equally. Specifically, universities that are endowed with extensive resources are less likely to be negatively affected by the increased need for specialized equipment. For instance, the introduction of Bitnet, an early version of the Internet, produced a greater benefit to mid-tier university researchers relative to elite university researchers (Agrawal & Goldfarb, 2008). The authors found that Bitnet reduced the collaboration costs, which, in turn, facilitated increased presence of mid-tier universities in research. Similarly, Ding, Levin, Stephan, & Winkler (2010) found that increased IT helped scientists from non-elite universities and female scientists more relative to elite universities and male scientists, respectively. Based on this information, we argue that non-elite universities would be negatively affected by the rise of deep learning.

In sum, other things being equal, increased importance of compute, data and AI talent would lower the KP for non-elite universities. In contrast, the KP would be higher for elite universities and large firms given their access to these inputs. Overall, we should observe a divergence between the two groups – have and have-nots – in modern AI research. In the next sections, we systematically explore whether and to what extent the rise of deep learning de-democratizes AI research.

**Empirical Strategy: The Generalized Synthetic Control Method**

We aim to estimate the causal impact of deep learning's sudden prominence on organizational participation in AI conferences. While the difference-in-difference is a widely used method to estimate causal impact, it requires a strong "parallel trend" assumption. As Figures 1 and 2 illustrate, we cannot confirm that the parallel trend condition has been fulfilled. Because we have panel data, we utilize a recently developed method (Xu, 2017) called the Generalized Synthetic Control Method (GSC Method) to directly impute counterfactuals for our treated units (AI conferences) from control units (other computer science conferences). This method is an improvement over the previously widely used synthetic control method (Abadie, Diamond, & Hainmueller, 2015). The premise of a synthetic control method is that a combination of untreated units may be a better comparison to the treatment unit(s) than a single untreated unit, which is often used in the difference-in-difference method. A synthetic control is a "weighted average" of similar untreated units that can replicate similar characteristics of the treated unit in the pre-intervention period. A synthetic control allows us to use that control's predicted post-intervention period as the counterfactual for the treated unit.

The GSC method has three distinct advantages over existing methods, including the original synthetic control method. First, this method relaxes many assumptions that linear two-way fixed effects models often violate, such as constant treatment effect and the absence of time-varying confounders. Second, the GSC method accommodates multiple treated units, which allows us to use all the AI conferences in this setting as treated units. Finally, using a parametric bootstrap procedure, the GSC method provides easily interpretable uncertainty estimates. We use the *gsynth* R package to estimate the average treatment effect on the treated (ATT).

Let's assume the outcome—the number of papers associated with a certain group of organizations—is $Y_{it}$ for conference i in period t. For each AI conference the treatment indicator $\delta_{it}$ takes 1 after the ImageNet shock. For non-AI conferences, this value takes 0 throughout the whole period.

The functional form of the outcome can be written as follows:

$$Y_{it}=\delta_{it}\,\theta_{it} + x'_{it}\,\beta + \lambda_i' f_t + \alpha_i + \eta_t + \varepsilon_{it} \qquad (2)$$

Here, $f_t$ is a vector of time-varying latent common factors, and $\lambda_i$ is a vector of factor loadings or conference-specific intercepts. The number of factors is selected using a cross-validation process. $x_{it}$ is a vector of observed covariates of the conferences (e.g., the total number of papers). $\beta$ is a vector of unknown parameters and $\varepsilon_{it}$ represents the error term; $\alpha_I$ and $\eta_t$ are individual and time fixed effects, respectively. $\theta_{it}$ is the treatment effect of $\delta_{it}$ on conference i in period t. Standard errors and confidence intervals are calculated using 2000 bootstraps blocked at the conference level.

To quantify the impact of ImageNet shock, for each conference, we need both the observed outcome $Y_{it}(1)$ and the counterfactual $Y_{it}(0)$ where the ImageNet shock did not happen. To estimate $Y_{it}(0)$ for each treated unit in the post-treatment period, the GSC method uses equation (2). Finally, the ATT is the mean difference between $Y_{it}(1)$ and $Y_{it}(0)$ in the post-treatment years.

**Identification Strategy**

To measure the causal impact of deep learning's unanticipated rise on large organizations' involvement in AI research, we rely on an exogenous shock to the broader computer science community. The sudden change in compute-usage since 2012 has been attributed to the success of deep learning in an academic contest known as the ImageNet contest. Since 2010, the ImageNet Large Scale Visual Recognition Challenge (ILSVRC) had been evaluating AI models for object detection and image classification at a large scale. Before 2012, deep learning models were "*too slow for large-scale applications*" (Raina, Madhavan, & Ng, 2009: 873) and there was no compelling evidence that they were suitable for large scale models. In 2012, at the ImageNet contest, one particular deep learning model known as AlexNet produced results that surprised many academics and industry experts. The organizer of the ImageNet contest termed deep learning's success an "unexpected outcome".[1] The winning team demonstrated that GPUs could be used to unleash neural networks' capabilities, which was surprising to most observers (Badawi et al., 2018; Krizhevsky, Sutskever, & Hinton, 2012). The reaction to the ImageNet's success surprised even the winner of that competition.[2] Since then, dozens of firms have acquired startups, increased R&D in AI, and filed for AI-related patents. The Economist summarized this phenomenon as follows: *"The rehabilitation of 'AI', and the current excitement about the field, can be traced back to 2012 and an online contest called the ImageNet Challenge*" (2016). Other computer scientists (Alom et al., 2018; Russakovsky et al., 2015) and journalists (Gershgorn, 2018) also supported this claim. Put simply, within computer science, the widespread use of GPUs for training deep learning was unanticipated, yet it allowed certain actors to take advantage of the research opportunity provided by the increased compute.

[1] Fei Fei Li's presentation on ImageNet: https://web.archive.org/web/20180131155559/http://image-net.org/challenges/talks_2017/imagenet_ilsvrc2017_v1.0.pdf (Page: 55)

In sum, due to the ImageNet 2012's surprising result, the broad use of GPUs as a training technology for deep learning was unanticipated within the AI community. As such, this setting provides a natural experiment to draw causal inference about observed increased participation of certain organizations triggered by the unanticipated availability of compute. Put differently, the unexpected GPU use for deep learning is an exogenous event that is correlated with increased availability of compute but not with research organizations' participation behavior or abilities. However, increased compute indirectly affected organizations' research activities and the tendency to exploit opportunities that were opened by the sudden availability of increased compute. This is why, in a later section, we use 2012 as the intervention year for our empirical analysis.

**Data**

We combine data from multiple sources: csrankings.org, Scopus, QS World University Rankings, the US News & World Report's Rankings, and Fortune magazine. Data was collected from Elsevier Scopus, which is one of the largest repositories of academic publications. Scopus has several advantages over other similar sources. First, this site focuses on peer-reviewed publications. Second, Scopus allows one to extract affiliations data and maintains multiple affiliations of an author, which Microsoft Academic Graph does not. This is important because a significant number of researchers within AI have dual affiliations (Gofman & Jin, 2019).

To collect computer science publication data, we consulted csrankings.org. This website was created by an academic computer scientist who made a list of the most prestigious publication venues to rank computer science department programs based on surveys and consultations with other leading computer scientists. It has been widely used in the literature (Gofman & Jin, 2019; Meho, 2019; Yang, Gkatzelis, & Stoyanovich, 2019). In computer science, conference publication plays a significant role in the tenure process (Freyne et al., 2010). Csrankings.org lists only the most prestigious conferences in each subfield of computer science, and the categorizations are based on the Association for Computing Machinery (ACM) Special Interest Groups (SIG). The curator argues, "*only the very top conferences in each area are listed. All conferences listed must be roughly equivalent in terms of number of submissions, selectivity and impact to avoid creating incentives to target less selective conferences.*"[3] Moreover, the conferences are comparable in submissions and selectivity; this helps create valid counterfactuals in the latter section.

[2] "[Alex] Krizhevsky, [..], chuckles when recalling the weeks after the 2012 ImageNet results came out. "It became kind of surreal," he says. "We started getting acquisition offers very quickly. Lots of emails." (Gershgorn, 2018)

Csranking.org allowed us to list 63 most prestigious conferences across all major areas in computer science including computer vision, machine learning and data mining, computer architecture, and mobile computing. In total, we collected 175,491 papers published from the years 2000 to 2019. From this dataset, we took a subset of conferences with at least 25 conference papers for a given year.[4] We excluded one AI conference, NAACL, which had fewer than 6 observations before 2012.[5] Our final sample includes 171,394 peer-reviewed articles from 57 conferences. The full conference list is available in Appendix A, Table A1. The panel structure of the data is presented in Appendix B, figure B1.

Based on the csranking.org's list we used 4 sub-areas of AI as a treated conference: *Artificial Intelligence, Computer Vision, Machine Learning & Data Mining, and Natural Language Processing*. Our consultation with computer scientists suggests that these 4 sub-areas have been affected by deep learning. We include 10 conferences under these 4 sub-areas of computer science: Association for the Advancement of Artificial Intelligence (AAAI), The International Joint Conference on Artificial Intelligence (IJCAI), Conference on Neural Information Processing Systems (NeurIPS), International Conference on Machine Learning (ICML), Conference on Knowledge Discovery and Data Mining (KDD), Association for Computational Linguistics (ACL), Empirical Methods in Natural Language Processing (EMNLP), Conference on Computer Vision and Pattern Recognition (CVPR), European Conference on Computer Vision (ECCV), and International Conference on Computer Vision (ICCV).

[3] http://csrankings.org/faq.html

[4] Fewer than 25 papers for a particular conference for a given year could indicate that Scopus did not have the full conference proceedings for that year. Out of 966 conferences, only 63 observations had between 1 and 24 papers. We ran additional models with at least 10 papers, our results are robust in this sample too.

[5] The GSC method requires at least 6 observations before the treatment to create a reliable counterfactual.

In the same way, we included non-AI conferences that were not affected by deep learning. The list includes all the top-tier conferences across all the major sub-areas of computer science such as the Annual International Conference on Mobile Computing and Networking (MobiCom) in Mobile Computing, Symposium on Foundations of Computer science (FOCS) in Algorithms and Complexity, and Conference on Human Factors in Computing Systems (CHI) in Human-Computer Interaction. The donor pool or control units come from these non-AI conferences. This is a reasonable donor pool because the same set of firms and universities are more likely to be active in the same set of computer science conferences. Moreover, the conferences are similar in terms of the number of submissions and selectivity.

We also collected data from the Fortune Global 500 list's 2018 edition. In particular, we focused on 46 technology companies on that list as classified by Fortune (the full list is presented in Appendix A, Table A2). For academic ranking, we collected data from the QS World University Ranking's 2018 edition and the US News and World Report Global University Ranking's 2018 version. These rankings have been widely used in the literature and have proven to influence administrative actions taken by university officials and students alike (Sauder, 2008).

**Variables**

Our unit of analysis is the conference. Our primary dependent variable is the total number of papers published by a specific group (e.g., firms or elite universities). Following the literature (Arora et al., 2018; Frank et al., 2019), we calculate the total number of papers that have at least one author who is affiliated with that specific group. To classify affiliations, at first, we used a fuzzy string matching and regular expressions to classify and label the data. Both authors and three research assistants reviewed all the unclassified observations to minimize misspelling related misclassification.

*Fortune500Tech* includes publications for 46 firms listed by Fortune magazine, which are presented in Table A2 of Appendix A. To label affiliation, we used the larger entities as corresponding institutions. For instance, DeepMind, Google AI, Waymo, Google Brain, have all been counted under Google. We counted only once under this variable even if the paper had multiple authors from different firms from Fortune500Tech to avoid double counting.[6] Similarly, we calculated *Fortune500NonTech*, which includes publications from other firms in the Fortune 500 list that are not labeled as technology firms. To create *Non-Fortune500firms*, we first listed major firms with a higher number of AI publications. We then looked for organizational names with an extensive list of keywords that includes 'ltd', 'llc', 'inc', 'limited', 'consult', 'industries', 'llp', 'gmbh', 'corp', 'incorporated', 'incorporation', 'corporation', and 'company". Both authors and three research assistants reviewed the list to cross-check misspelling and multiple variations of firm names. All of these firms were counted under *Non-Fortune500firms*. To calculate *AllFirms*, we included all the firms that are listed under Fortune500, as well as firms not listed in Fortune 500; we also used the extensive keywords list to include non-prominent firms. The raw data for the *AllFirms* variable is presented in Figure B2 in Appendix B.

[6] Our estimates from these variables are *conservative* estimates. However, in the robustness analysis we took another weighted approach to count the co-authorships. Our estimates are presented in Table 5.

For academic institutions too, we counted the large entities. For instance, all the institutions under MIT, such as MIT Computer Science & Artificial Intelligence Laboratory, MIT Department of Electrical Engineering and Computer Science, and MIT Media Lab were labeled as MIT. We conducted extensive manual checking to ensure that different variations of a lab name, including abbreviations, were attended to in the process of including them under the larger entities. As before, while creating QS1-50, we counted only once even if two universities within QS1-50 collaborated. In the same way, we calculated *QS51-100, QS101-200, QS201-300, QS301-500.*

Our independent variable is *ImageNet2012*, a binary variable that indicates whether a conference is treated or not. For AI conferences from 2012, this variable takes the value 1, and otherwise, it takes the value 0. As a control variable, we include *TotalNumOfPaper*, which captures the total number of papers of a conference in a given year. The total number of papers controls for the popularity and selectivity of a conference.

**Summary Statistics**

First, we present the descriptive statistics in Table 1.

**Table 1: Summary Statistics**

| Statistic | N | Mean | St. Dev. | Min | Max |
|---|---|---|---|---|---|
| TotalNumofPaper | 903 | 189.45 | 238.79 | 25 | 1,648 |
| ImageNet2012 | 903 | 0.08 | 0.27 | 0 | 1 |
| QS1-50 | 903 | 66.90 | 87.43 | 3 | 762 |
| QS51-100 | 903 | 35.42 | 44.09 | 0 | 310 |
| QS101-200 | 903 | 32.05 | 42.29 | 0 | 347 |
| QS201-300 | 903 | 20.02 | 26.38 | 0 | 220 |
| QS301-500 | 903 | 22.22 | 31.72 | 0 | 205 |
| Allfirms | 903 | 41.68 | 57.90 | 0 | 680 |
| Fortune500Tech | 903 | 23.22 | 34.32 | 0 | 384 |
| Fortune500NonTech | 903 | 6.39 | 10.21 | 0 | 71 |
| Non-Fortune500firms | 903 | 14.28 | 22.73 | 0 | 309 |
| Historically Black College and Universities(HBCU) | 903 | 0.71 | 1.54 | 0 | 16 |
| Hispanic Association of College and Universities(HACU) | 903 | 6.26 | 7.99 | 0 | 60 |
| Compute | 20 | 10,981.45 | 41,700.14 | $6 \times 10^{-8}$ | 186,000 |
| **Collaboration** | | | | | |
| Firm's-Only pub | 903 | 9.48 | 12.35 | 0 | 113 |
| Firm-University Collaboration | 903 | 27.75 | 42.04 | 0 | 522 |
| QS1-50 & Firm Collaboration | 903 | 13.47 | 22.39 | 0 | 273 |
| QS51-100 & Firm Collaboration | 903 | 6.97 | 10.64 | 0 | 129 |
| QS301-500 & Firm Collaboration | 903 | 3.39 | 5.92 | 0 | 77 |
| QS1-50 & FortuneGlobal500Tech Collaboration | 903 | 8.63 | 14.92 | 0 | 176 |
| QS51-100 & FortuneGlobal500Tech Collaboration | 903 | 4.59 | 7.31 | 0 | 85 |
| QS301-500 & FortuneGlobal500Tech Collaboration | 903 | 1.96 | 3.65 | 0 | 46 |

Note: This table presents the summary statistics for 57 AI and non-AI conferences. The unit of observation is conference-year. The data are an imbalanced panel for 903 observations from 2000 to 2019.

The descriptive table reveals that there is a strong correlation between ranking and average publications. For instance, elite universities have a strong presence in broader computer science with 66 average papers per year per conference. However, universities ranked from 51 to 100 publish 35 papers, whereas 100 universities ranked between 101-200 publish only 32 papers. Moreover, on average, only 22 papers are affiliated with the QS 301-500 ranked universities. Further, *Allfirms* have a noticeable presence with almost 42 average papers per conference each year and large firms have an annual presence with 23 papers per conference.

**The Underrepresentation of Black and Hispanic-serving Institutions in AI**

To understand the participation of Historically Black Colleges and Universities (HBCU) in the AI conferences, we compiled a list of Historically Black Colleges and Universities from the US NEWS 2018 Best Colleges rankings. The list consists of 55 universities, 24 of which have at least one publication in AI conferences in the period from 2000 to 2019. All the HBCU institutions together have only 260 publications in AI conferences in that period, and the distribution is heavily skewed. For example, among them, South Carolina State University alone has 157 publications. Similarly, we compile the list of Hispanic-Serving Institutions from the Hispanic Association of College and Universities (HACU) from the association's website, leaving out the community colleges.[7] The list consists of 278 institutes, 49 of them have at least one publication in AI conferences in the period between 2000 and 2019. In total, they contributed to 2913 publications in top AI conferences in that period. For comparison, Microsoft alone has 3302 publications in top AI conferences in that period, which is more than the combined total AI publications of all HBCU- and HACU-affiliated institutions.

[7] https://www.hacu.net/

**The Increased Presence of Firms in AI research**

To illustrate an intuitive overview of the firms' presence, we present the share of papers affiliated with firms over time at AI conferences in figure 1. Figure 1 reports the annual share of all firms at top AI conferences. The graphs highlight a meaningful shift in corporate presence in AI research. This increased presence of corporations is more pronounced over the last few years. Figure 1 suggests that all ten conferences have experienced an upward trend in corporate representation.

**Figure 1: Firms' share of papers in major AI conferences**

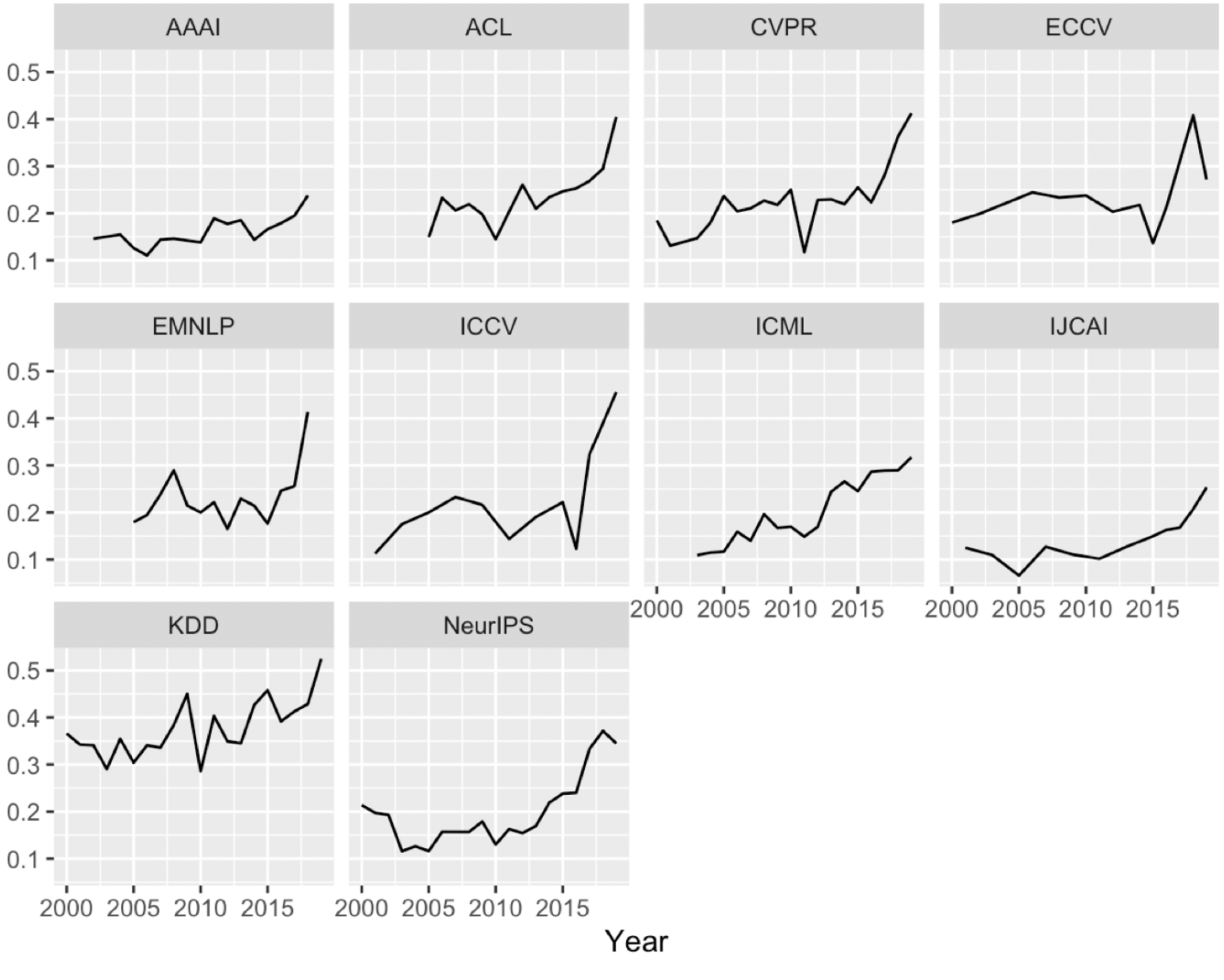

Note: This figure illustrates the share of papers that have at least one firm-affiliated co-author. For instance, a 0.30 value indicates that 30% of the papers at that conference in a particular year have at least one co-author from a firm.

Figure 2 illustrates the share of non-AI computer science conferences. Firms' publications in non-AI conferences within computer science do not display a consistent pattern as they do in AI conferences. In most cases, the level of firm publications is relatively stable. Another noticeable trend is that on average, the corporate participation rate across both AI and non-AI conferences was similar before 2012. Only after the 2012's ImageNet shock is an increase in firm-participation in AI noted.

**Figure 2: Firms' share of papers in major non-AI conferences**

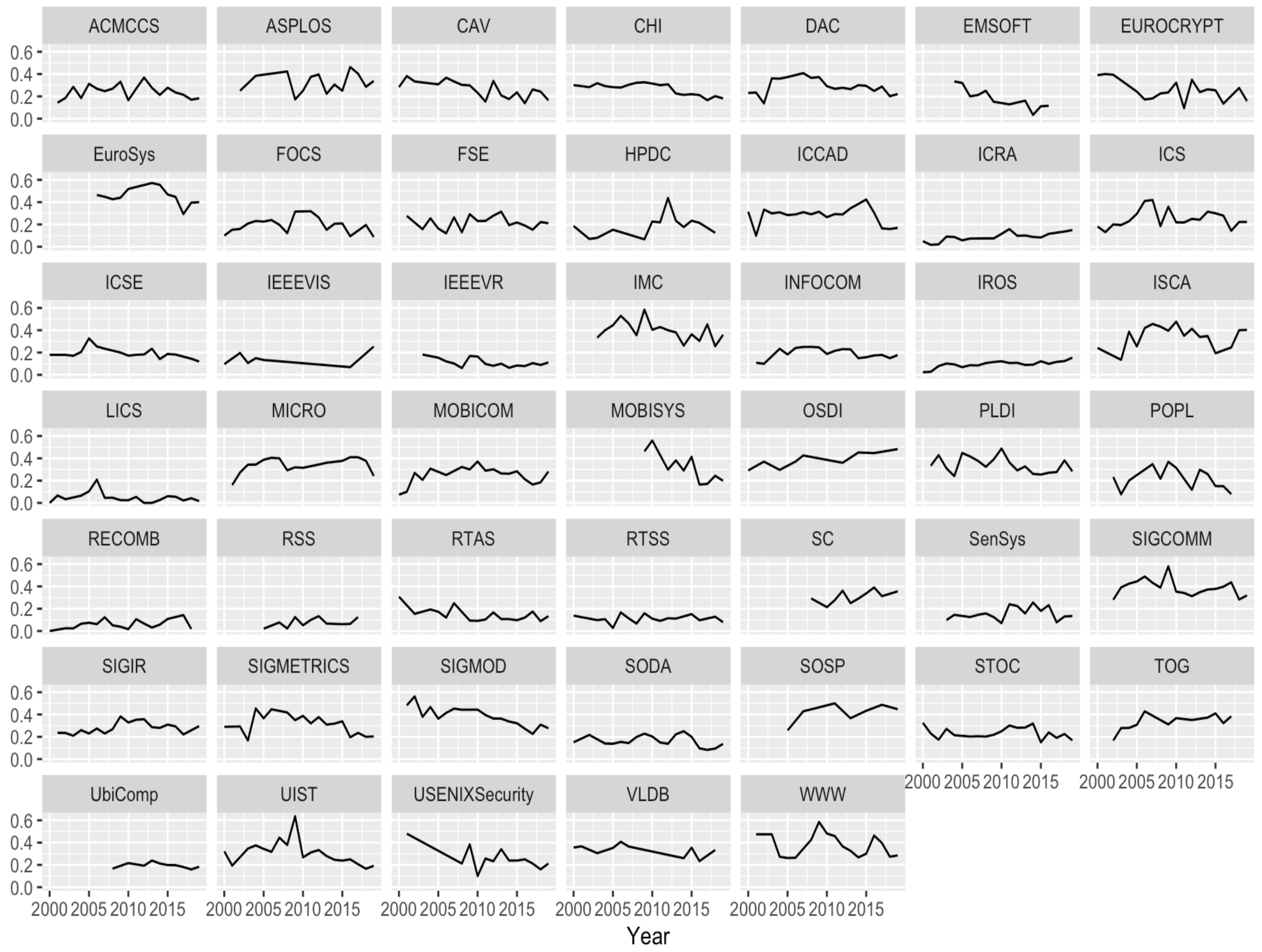

Note: This figure illustrates the share of papers that have at least one firm-affiliated co-author. For instance, a 0.20 value indicates that 20% of the papers at that conference in a particular year have at least one co-author from a firm.

## Results

Table 2 shows estimates of the effect of ImageNet's 2012 shock on the firm-level participation in AI conferences using the GSC method. This table presents the average treatment effect on the treated (ATT). All models include conference fixed effects, which account for heterogeneity in the underlying quality and popularity of individual conferences and year fixed effects, which control for conference-invariant changes over time. Table 2, Model 1 indicates that all firms have increased annual publication by 46 papers per conference. More specifically, Model 2 highlights that this increased corporate presence is primarily driven by Fortune500Tech firms, where only 46 firms increased their presence by more than 44 papers. Over 8 years, this amounts to more than 350 additional papers than the counterfactual. In other words, large technology firms are publishing 350 additional papers over an 8-year period per conference due to the rise of deep learning. However, Model 3 indicates that other large non-technology firms, namely Fortune500*Non*Tech, did not have any discernible impact in AI research. These results are consistent with recent research, which suggests that the accumulation of AI intangible capital is concentrated within specific industries and firms (Tambe, Hitt, Rock, & Brynjolfsson, 2019). In contrast, Model 4 indicates that firms not listed in Fortune500 have also increased their presence. More specifically, this increased presence is due to technology firms like Baidu, Nvidia, Uber, and SenseTime, which were not listed in Fortune500 Tech's 2018 edition. Interestingly, all of these firms have access to compute and have hired talented scientists.

One concern could be that the increased presence of firms is due to only a few large firms like Google and Microsoft. To counter that concern, we removed the 10 largest technology firms as listed by Forbes (2018)—Apple, Samsung, Microsoft, Google, Intel, IBM, Facebook, Tencent, Foxconn, and Oracle—from the variable *AllFirms*. Results are presented in Model 5, which shows that even after excluding the largest 10 technology firms, we observe an increased presence of firms in top AI venues. However, in this case, firms are increasing their presence by only 20 papers or less than half of Model 1. In other words, large technology firms seem to be playing an outsized role in increased industry presence. Overall, our results suggest that firms have increased participation in AI research; however, this increased presence is primarily driven by large technology firms. In sum, Models 1-5 are consistent with the argument that compute plays an important role because large technology firms have access to higher computing power.

**Table 2: The effects of ImageNet shock on firm-level participation in AI: GSC method estimates**

| | *Dependent variable:* | | | | |
|---|---|---|---|---|---|
| | (1) | (2) | (3) | (4) | (5) |
| | AllFirms | Fortune500Tech | Fortune500NonTech | Non-Fortune500firms | AllFirms(Excluding top10) |
| ImageNet2012 | 46.63*** | 44.50*** | -0.910 | 16.13*** | 19.94*** |
| | (5.905) | (2.208) | (2.071) | (1.345) | (2.294) |
| TotalNumOfPaper | 0.185*** | 0.069*** | 0.047*** | 0.066*** | 0.108*** |
| | (0.006) | (0.005) | (0.004) | (0.002) | (0.004) |
| Conference Fixed Effects | Yes | Yes | Yes | Yes | Yes |
| Year Fixed Effects | Yes | Yes | Yes | Yes | Yes |
| Observations | 903 | 903 | 903 | 903 | 903 |
| Treated Conference | 10 | 10 | 10 | 10 | 10 |
| Control Conference | 47 | 47 | 47 | 47 | 47 |

Note: This table estimates the 2012's ImageNet shock on firms' participation in major 10 AI conferences. We used 47 non-AI conferences as control groups. Standard errors and 95% confidence intervals are computed using 2,000 parametric bootstraps blocked at the conference level. Standard errors are shown in parentheses and *, **, and *** denote significance at the 10%, 5%, and 1% level, respectively.

Next, we turn to large technology firms or *Fortune500Tech* firms. We graphically explore the ImageNet shock's dynamic treatment effect. Figure 3 illustrates the average number of *Fortune500Tech* firm publications (solid line) and the average predicted number of large technology firms' publications (dashed line) in the absence of a deep learning breakthrough or the counterfactual. This figure shows that before 2012 (indicated with 0), the two lines are well matched; the better the fit in previous years, the more reliable the model is. After 2012, we observe a noticeable change between the observed number of publications and the counterfactual's number of publications. This demonstrates that the sudden rise of deep learning resulted in large technology firms' presence increasing significantly. The effect of ImageNet rises gradually over time; in particular, the effect size is more salient since 2016. This is consistent with OpenAI's observation on compute use within modern AI research (Amodei & Hernandez, 2018). OpenAI's research suggests that until 2014, training on GPU was relatively uncommon, which is consistent with Figure 3's illustration that the treatment effect was much smaller from 2012 to 2014. In line with the availability of compute argument, a higher number of GPU (10-100) usage became common between 2014 and 2016, which allowed a significant improvement in deep learning. However, OpenAI posits that greater algorithmic parallelism with specialized hardware such as Tensor Processing Units or TPUs became available only from 2016. Figure 3 also highlights that ATT is much higher over the last few years.

**Figure 3: The effects of ImageNet shock on Fortune500Tech firms' participation in AI**

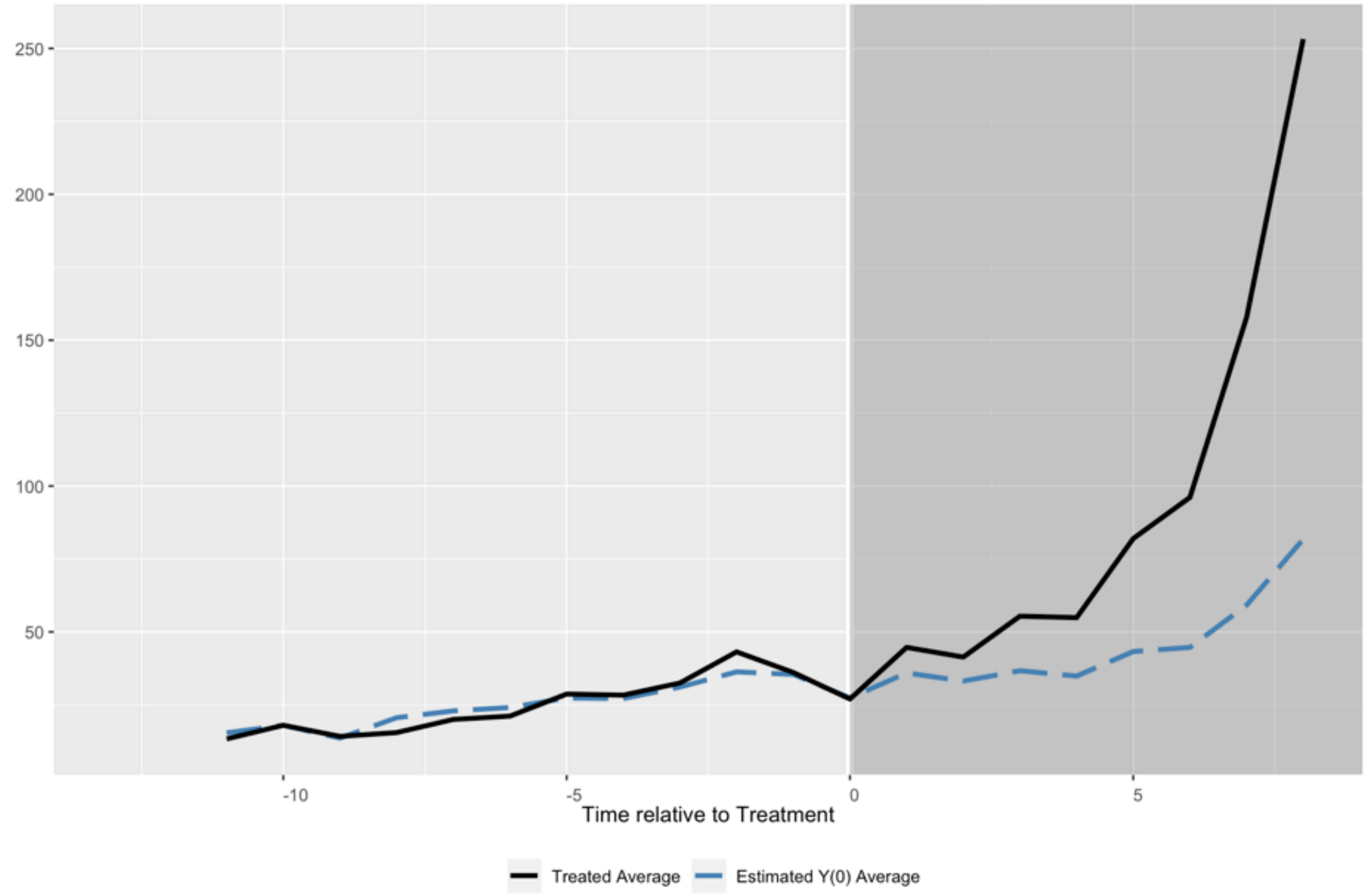

Note: This figure illustrates the dynamics of the estimated ATT. The black line represents the average number of annual publications by firms at top AI conferences and the blue dotted line represents the predicted average number of publications. In the post-treatment period, a wider gap indicates a higher treatment effect. The shaded area denotes the treated periods.

Figure 4 illustrates the gaps between the actual number of publications and the predicted number of publications or the counterfactual. Put simply, this figure presents the average treatment on the treated or ATT for Fortune500Tech firms. In the pre-treatment periods, before 2012, ATT was close to zero. However, ATT takes off right after the ImageNet shock and reaches around 180 publications in 2019.

**Figure 4: Average treatment on the treated (ATT) of the ImageNet shock on Fortune500Tech firms' participation**

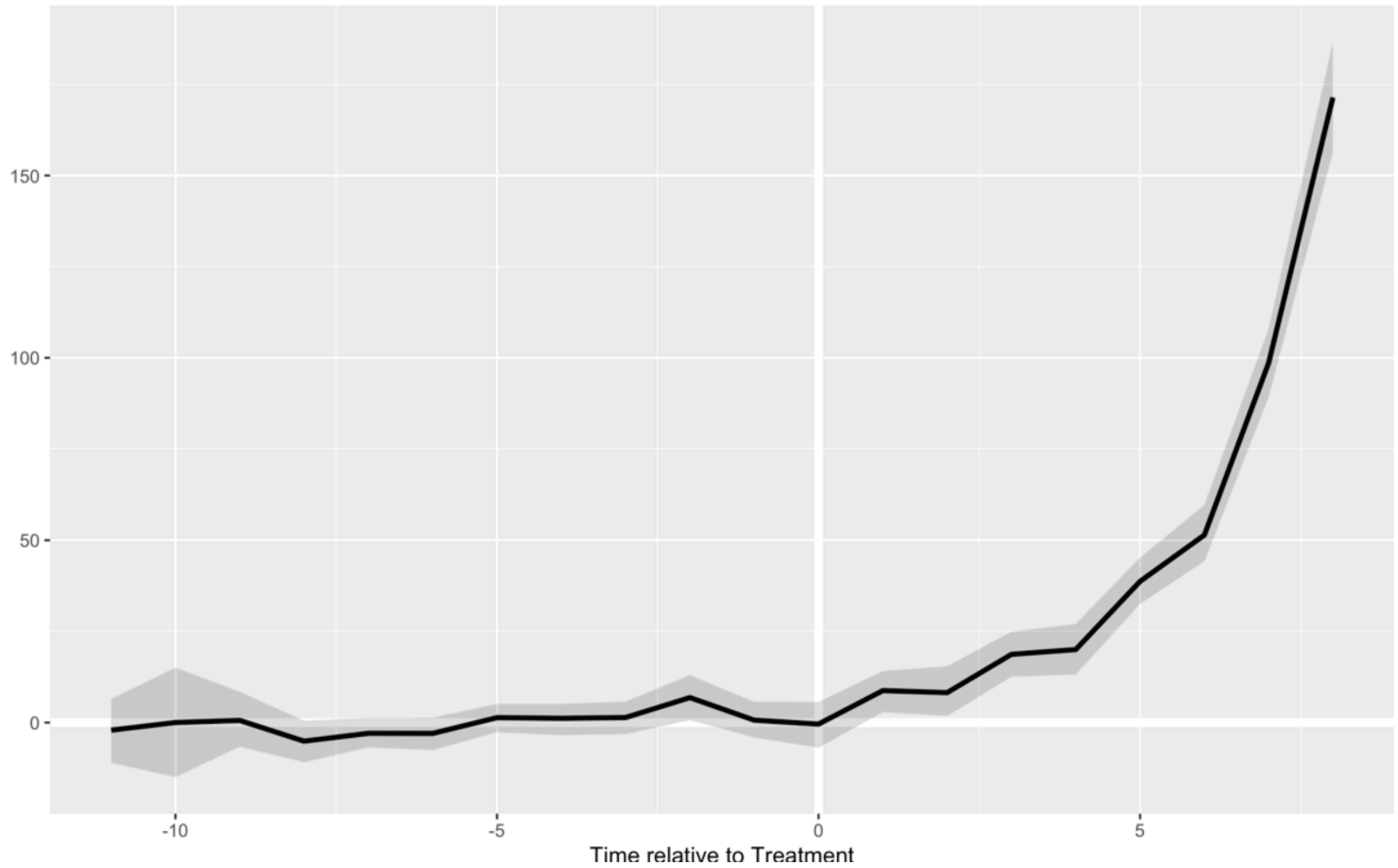

Note: This figure illustrates the dynamic treatment effects estimates from the GSC method. The dark line indicates the estimated average treatment effect on the treated (ATT) and 95% confidence interval. Standard errors and 95% confidence intervals are computed using 2,000 parametric bootstraps blocked at the conference level.

In sum, Figures 3 and 4 both highlight the significant impact of the 2012 ImageNet shock on firm-level participation in AI research.

**Firms' Collaboration Strategy**

Next, we delve deeper into how firms have been able to increase their presence. We document two strategies that helped firms increase their presence in AI research. First, firms increased firm-only publications or publications that did not have any outsider collaborator. Interestingly, *firm-only* publication has increased by only 8 publications annually per conference, as presented in Table 3, Model 1. Second, firms increased publications jointly produced with universities by almost 42 additional papers or five times more than firm-only publications. However, we find that firms mostly increased collaborations with elite universities. The results from Models 3 and 5 suggest that firms are collaborating *six times more* with QS ranked top 50 universities than QS ranked 301-500 universities combined. This is consistent with recent research, which documents that academic funding is increasingly "over attracted" by elite universities that tend to collaborate with each other (Szell & Sinatra, 2015). Models 3 and 6 suggest that this collaboration is primarily driven by large technology firms. In other words, large technology firms are mostly responsible for the majority of the collaborations.

One potential reason behind these collaborations between elite universities and large technology firms could be resource complementarity. On the one hand, large technology firms have compute and large proprietary datasets. On the other hand, elite universities have AI scientists who have the tacit knowledge in deep learning. Therefore, collaboration between these two groups creates a win-win situation. However, this increased collaboration might have inadvertently crowded out mid-tier and lower-tier universities from these competitive conferences.

**Table 3: The Collaboration Strategies of firms: GSC method estimates**

| | Collaboration | | | | | | | |
|---|---|---|---|---|---|---|---|---|
| | (1) | (2) | (3) | (4) | (5) | (6) | (7) | (8) |
| | | | (All Firms) | | | (Fortune500 Tech) | | |
| | Firms-Only | Firm-University | QS1-50 | QS51-100 | QS301-500 | QS1-50 | QS51-100 | QS301-500 |
| ImageNet2012 | 8.24*** | 41.42*** | 26.15*** | 7.98*** | 4.10*** | 20.74** | 7.95*** | 4.15*** |
| | (1.57) | (2.04) | (1.25) | (0.80) | (0.47) | (.91) | (0.66) | (0.34) |
| TotalPapers | 0.026*** | 0.123*** | 0.047*** | 0.032*** | 0.020*** | 0.025*** | 0.015*** | 0.008*** |
| | (0.002) | (0.004) | (0.002) | (0.002) | (0.000) | (0.002) | (0.001) | (0.000) |
| Conference FE | Yes | Yes | Yes | Yes | Yes | Yes | Yes | Yes |
| Year FE | Yes | Yes | Yes | Yes | Yes | Yes | Yes | Yes |
| Observations | 903 | 903 | 903 | 903 | 903 | 903 | 903 | 903 |
| Treated Conference | 10 | 10 | 10 | 10 | 10 | 10 | 10 | 10 |
| Treated Conference | 47 | 47 | 47 | 47 | 47 | 47 | 47 | 47 |

Note: This table estimates the 2012's ImageNet shock on firms' collaboration strategies in AI conferences. We used 10 AI conferences as the treatment group and 47 non-AI conferences as the control group. Standard errors and 95% confidence intervals are computed using 2,000 parametric bootstraps blocked at the conference level. Standard errors are shown in parentheses and *, **, and *** denote significance at the 10%, 5%, and 1% level, respectively.

**The Differential Effects of ImageNet Shock on University Participation in AI Research**

Having shown that the rise of deep learning resulted in the increased presence of firms in AI research, we now turn our attention to universities. We repeat the same process using the GSC method with conference and year fixed effects. Table 4, Model 1 indicates that the top 50 universities in the QS ranking have increased presence by more than 40 papers. This is a significant increase given that on average these elite universities publish around 67 papers annually per conference. Aggregated over eight years, the increased presence in AI publications becomes substantial: more than 320 additional papers per year per conference. Models 2 and 3 show that the QS News ranking 51-100 ranked universities and the QS News ranking 101-200 ranked universities have not observed any significant impact resulting from the ImageNet shock. However, Model 4 reports that mid-tier universities such as 201-300 ranked universities have been publishing 8 fewer articles in top AI conferences than their counterfactuals. Similarly, universities that are ranked 301-500, together, are publishing almost 6 fewer papers per year per conference. This is particularly significant given that these universities publish on average 22 papers per year per conference. In other words, they are publishing, on average 25% fewer papers than the counterfactual since the rise of deep learning. Model 6 presents the result for HBCU universities. The result suggests that the rise of deep learning had no discernible impact on HBCU participation. One potential explanation could be their already low presence in AI research.

These findings suggest that the divergence between elite universities and unranked universities grows with the ranking disparity. Taken together, our findings offer support for the proposition that while large firms and elite universities are gaining ground in AI research, their increased presence is crowding out mid-tier and lower-tier universities. This result is particularly troubling because in other academic areas such as life sciences and economics, non-elite universities are catching up in research with elite universities (Halffman & Leydesdorff, 2010; Kim et al., 2009).

**Table 4: The differential effects of ImageNet shock on university participation in AI: GSC method estimates**

| | *Dependent variable:* | | | | | |
|---|---|---|---|---|---|---|
| | (1) | (2) | (3) | (4) | (5) | (6) |
| | QS1-50 | QS51-100 | QS101-200 | QS201-300 | QS301-500 | HBCU |
| ImageNet2012 | 40.18*** | 4.19 | -0.09 | -14.34** | -5.57*** | -0.633 |
| | (4.49) | (6.68) | (5.64) | (5.14) | (1.54) | (0.596) |
| TotalNumOfPaper | 0.282*** | 0.175*** | 0.179*** | 0.126*** | 0.155*** | 0.002*** |
| | (0.008) | (0.011) | (0.010) | (0.005) | (0.003) | (0.000) |
| Conference FE | Yes | Yes | Yes | Yes | Yes | Yes |
| Year FE | Yes | Yes | Yes | Yes | Yes | Yes |
| Observations | 903 | 903 | 903 | 903 | 903 | 903 |
| Treated Conference | 10 | 10 | 10 | 10 | 10 | 10 |
| Control Conference | 47 | 47 | 47 | 47 | 47 | 47 |

Note: This table estimates the 2012's ImageNet shock on the heterogeneity of university participation in AI conferences. We used 10 AI conferences as the treatment group and 47 non-AI conferences as the control group. Standard errors and 95% confidence intervals are computed using 2,000 parametric bootstraps blocked at the conference level. Standard errors are shown in parentheses and *, ** and *** denotes significance at the 10%, 5%, and 1% level, respectively.

**Placebo Tests**

To evaluate the credibility of our estimates, we use another subfield—computer science (Software Engineering)—as a placebo instead of the AI conferences. In other words, we can estimate the ATT using the GSC method for a conference that did not experience the ImageNet shock. A discernible placebo treatment effect would undermine confidence in our estimates. Here, we used two major conferences from the subfield *Software Engineering*: Conference on the Foundations of Software Engineering (FSE), and the International Conference on Software Engineering (ICSE). We repeat the same exercise as before using the GSC method. We present the dynamic treatment effect in Figure 5, which closely fits the observed value before 2012.

Figure 5 suggests that there is no significant treatment effect since 2012. In other words, in contrast to our original result (as in Figure 3), the sudden rise of deep learning in 2012 had no discernible effect on large technology firms' participation in the "Software Engineering" subfield. This observation increases confidence in our argument that the result is indeed due to the sudden rise of deep learning and not because of a limitation of our estimator.[8]

**Figure 5: Placebo test with "Software Engineering" subfield: The effects of ImageNet shock on Fortune500Tech firm's participation**

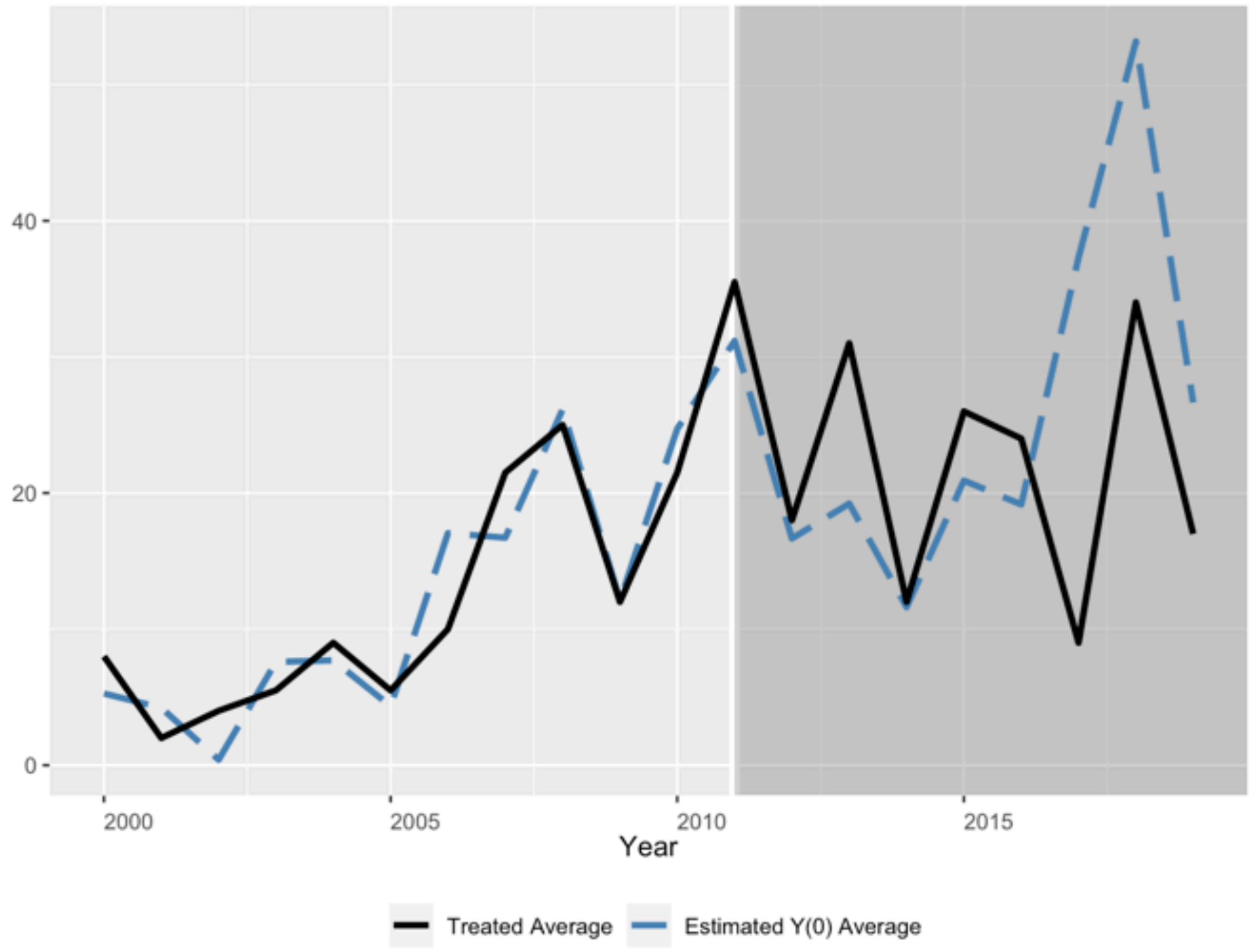

Note: Placebo counterfactual analysis with another computer science subfield (two non-AI conferences from "Software Engineering": FSE, ICSE). The black line represents the average number of annual publications by Fortune500Tech firms at top AI conferences and the blue dotted line represents the counterfactual.

To consider the role of time-varying confounding factors in the pre-treatment period, we conducted another placebo test following Liu et al. (2020). Instead of 2012, we introduced the ImageNet shock 3 years earlier in 2009. There should not be any discernible treatment effect of the ImageNet in this time period if there are no time-varying confounders. An ATT estimate which is statistically insignificant would increase confidence in our result. We use the *fect* R package to test the placebo in time. This test is robust to model misspecifications and relies on out of sample predictions to reduce overfitting.

[8] We ran the same placebo test with other non-AI computer science subfields; all the placebo analysis produced similar results.

**Figure 6: Placebo ImageNet shock in 2009: Fortune500Tech firms' participation in AI**

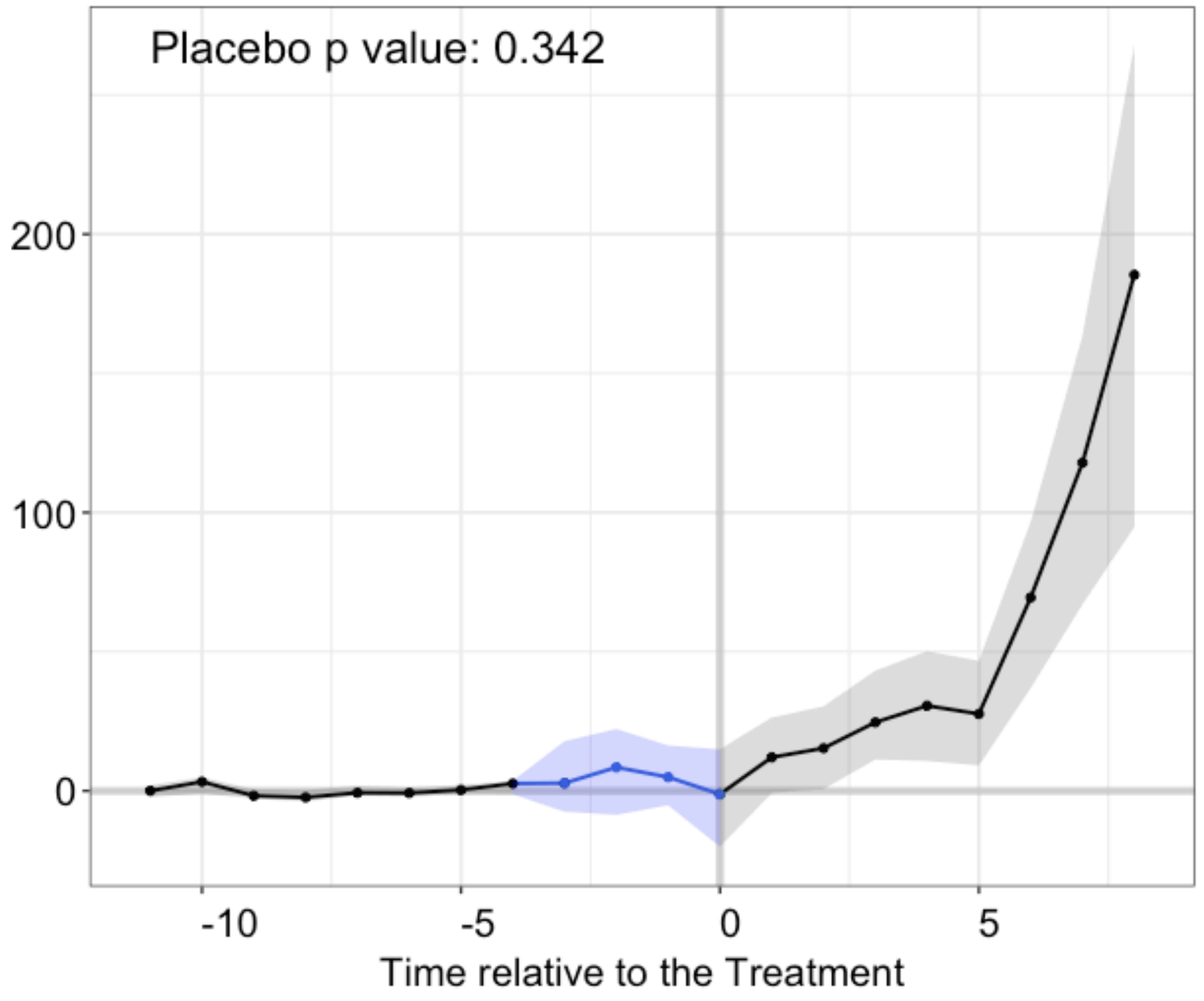

Note: This figure illustrates the placebo test with a different time (2009) as the intervention year. Standard errors and 95% confidence intervals are computed using 2,000 parametric bootstraps blocked at the conference level.

Our placebo test assumes that the ImageNet shock happened in 2009 rather than 2012. Indeed, Figure 6 illustrates no discernible treatment effect between 2009 and 2012 (the p-value is much higher than 0.05). This validates our argument that 2012's ImageNet challenge is the event that changed the AI research field significantly. In addition to this 3-year test, we also ran 2- and 4-years placebo tests. Overall, the results are similar and increase confidence in our estimates.

**Additional Robustness Tests**

For further robustness tests, we used a different method to count organizational participation. Instead of adding one to each group for each paper (which is widely practiced in the literature), we weighted the affiliations of the authors. For each paper, we assigned (1/total number of co-authors) * (1/total number of affiliations of an author) for each group (QS1-50, Fortune500Tech). This ensures that authors' multiple affiliations and the number of authors are weighted accordingly. In computer science, it is increasingly common for well-known scientists to have multiple affiliations (Recht, Forsyth, & Efros, 2018). For example, Yann LeCun, the Turing award winning scientist who is also known for his pioneering work on deep learning, has two institutional affiliations. At the same time, he is affiliated with New York

University and Facebook. If LeCun co-authors a paper with another author, each group (*QS51-100*, for New York University, and *Fortune500Tech* for Facebook) will get ½* ½ or ¼. If LeCun's co-author is only affiliated with Google, then *Fortune500Tech* will get ½* 1= ½.[9] For this analysis, we used the QS 2018 rankings. The results are presented in Table 5.

**Table 5: The effects of ImageNet Shock on organizational participation in AI: weighted-affiliation measure (QS 2018 ranking)**

| | *Dependent variable:* | | | | | | |
|---|---|---|---|---|---|---|---|
| | (1) | (2) | (3) | (4) | (5) | (6) | (7) |
| | AllFirms | Fortune500Tech | QS1-50 | QS51-100 | QS101-200 | QS201-300 | QS301-500 |
| ImageNet2012 | 26.04*** | 22.1*** | 18.27*** | -6.07*** | -6.18** | -11.94*** | -7.127*** |
| | (8.043) | (1.227) | (3.539) | (1.433) | (1.433) | (1.2) | (0.83) |
| TotalNumOfPaper | 0.079*** | 0.032*** | 0.183*** | 0.114*** | 0.113*** | 0.087*** | 0.093*** |
| | (0.004) | (0.002) | (0.005) | (0.003) | (0.006) | (0.002) | (0.001) |
| Conference FE | Yes | Yes | Yes | Yes | Yes | Yes | Yes |
| Year FE | Yes | Yes | Yes | Yes | Yes | Yes | Yes |
| Observations | 903 | 903 | 903 | 903 | 903 | 903 | 903 |
| Treated Conference | 10 | 10 | 10 | 10 | 10 | 10 | 10 |
| Control Conference | 47 | 47 | 47 | 47 | 47 | 47 | 47 |

Note: This table estimates the 2012's ImageNet shock on the heterogeneity of university participation in AI conferences using the weighted affiliated measurement. We used 10 AI conferences as the treatment group and 47 non-AI conferences as the control group. Standard errors and 95% confidence intervals are computed using 2,000 parametric bootstraps blocked at the conference level. Standard errors are shown in parentheses and *, **, and *** denote significance at the 10%, 5%, and 1% level, respectively.

The results in Table 5 are consistent with our previous results. Table 5, Model 1 shows that firms have increased presence since 2012. Similarly, Models 2 and 3 illustrate that large technology firms (Fortune500Tech) and elite universities (QS1-50) have increased their presence significantly. Taken together these three models estimate slightly lower than our previous estimates. This indicates a higher number of collaborations between elite universities and large firms. However, results from the weighted affiliation data are even more concerning for non-elite universities. For instance, Model 4's results are particularly notable because our previous results for QS51-100 did not have a negative significant impact. Otherwise, all the other non-elite universities report similar and negative significant impact of the ImageNet shock. Overall, our result is robust to different measurements of author affiliations.

One potential limitation of the estimates could be due to the idiosyncrasies related to the classification of elite and non-elite universities. To increase further confidence in our result, we performed additional analysis with the US News Global Ranking's 2018 edition instead of the QS Global Ranking. The results are reported in Table 6. Model 1 illustrates that indeed, elite universities have increased presence by 19 additional papers relative to the counterfactual.

[9] The total sum of these scores will be 1. For this particular paper, *Fortune500Tech* =¼ + ½, *QS51-100*= ¼

Similarly, Model 2 also reports that universities that are ranked from 51-100 in the US News have increased presence by 6 additional papers. On the other hand, Models 3,4 and 5 indicate that non-elite universities are struggling since 2012. More specifically, universities that are ranked from 201-300 are publishing almost 15 fewer papers than the counterfactual, which is significant at the 1% level. In other words, mid-tier universities are publishing almost 120 fewer papers over 8 years due to the sudden rise of deep learning. Overall, the results from the US News Global Ranking's 2018 edition are similar to the QS Ranking's 2018 edition.

**Table 6: The differential effects of ImageNet shock on university participation in AI: GSC method estimates (US News 2018 ranking)**

| | *Dependent variable:* | | | | |
|---|---|---|---|---|---|
| | (1) | (2) | (3) | (4) | (5) |
| | US1-50 | US51-100 | US101-200 | US201-300 | US301-500 |
| ImageNet2012 | 19.02*** | 6.40** | -7.137 | -14.85*** | -4.16 |
| | (4.099) | (2.498) | (8.657) | (2.199) | (3.251) |
| TotalNumOfPaper | 0.268*** | 0.195*** | 0.202*** | 0.132*** | 0.172*** |
| | (0.009) | (0.004) | (0.007) | (0.004) | (0.006) |
| Conference FE | Yes | Yes | Yes | Yes | Yes |
| Year FE | Yes | Yes | Yes | Yes | Yes |
| Observations | 903 | 903 | 903 | 903 | 903 |
| Treated Conference | 10 | 10 | 10 | 10 | 10 |
| Control Conference | 47 | 47 | 47 | 47 | 47 |

Note: This table estimates the 2012's ImageNet shock on the heterogeneity of university participation in AI conferences. We used 10 AI conferences as the treatment group and 47 non-AI conferences as the control group. Standard errors and 95% confidence intervals are computed using 2,000 parametric bootstraps blocked at the conference level. Standard errors are shown in parentheses and *, **, and *** denote significance at the 10%, 5%, and 1% level, respectively.

Another limitation could arise because ImageNet shock could potentially affect *both* university rankings and research behavior. To avoid this problem, we used the QS Ranking's 2011 edition instead of the 2018 ranking. The 2011 ranking would not be affected by the ImageNet shock because the shock happened in 2012. The results of this test are reported in Table 7 and are consistent with our previous estimates. Table 7, Model 1 indicates that elite universities have published 37 additional papers than the counterfactual, which is much higher than our both QS 2018 and US News 2018 estimates. In other words, universities that were elites in 2011 gained much more from the ImageNet shock. Similarly, Models 2 and 3 show that universities ranked between 51-100 and 101-200 did not experience any discernible impact. As seen in our previous estimates, universities ranked between 201 and 300 experienced noticeable negative impact. These universities published approximately 14 fewer papers than the counterfactual per year per conference. Finally, universities ranked between 301 and 500 also experienced a significant negative impact, publishing 6 fewer papers than the counterfactual, much like QS 2018's estimates.

**Table 7: The differential effects of ImageNet shock on university participation in AI: GSC method estimates (QS 2011 ranking)**

| | *Dependent variable:* | | | | |
|---|---|---|---|---|---|
| | (1) | (2) | (3) | (4) | (5) |
| | QS1-50 | QS51-100 | QS101-200 | QS201-300 | QS301-500 |
| ImageNet2012 | 36.92*** | -6.10 | -5.14 | -13.94*** | -6.40*** |
| | (4.05) | (3.93) | (4.79) | (5.13) | (1.57) |
| TotalNumOfPaper | 0.270*** | 0.191*** | 0.208*** | 0.130*** | 0.133*** |
| | (0.008) | (0.007) | (0.009) | (0.007) | (0.003) |
| Conference FE | Yes | Yes | Yes | Yes | Yes |
| Year FE | Yes | Yes | Yes | Yes | Yes |
| Observations | 903 | 903 | 903 | 903 | 903 |
| Treated Conference | 10 | 10 | 10 | 10 | 10 |
| Control Conference | 47 | 47 | 47 | 47 | 47 |

Note: This table estimates the 2012's ImageNet shock on the heterogeneity of university participation in AI conferences. We used 10 AI conferences as the treatment group and 47 non-AI conferences as the control group. Standard errors and 95% confidence intervals are computed using 2,000 parametric bootstraps blocked at the conference level. Standard errors are shown in parentheses and *, **, and *** denote significance at the 10%, 5%, and 1% level, respectively.

Taken together, Tables 6 and 7 suggest that the heterogeneous impact of the rise of deep learning on university participation is robust to different rankings and years.

For further robustness, we used another counterfactual estimator, the matrix completion estimator (Athey, Bayati, Doudchenko, Imbens, & Khosravi, 2018), which has been inspired by machine learning literature (Mazumder, Hastie, & Tibshirani, 2010). Using data from non-treated units, this method aims to construct a lower-rank representation of the outcome matrix. Interestingly, the GSC method has been demonstrated to be a special case of the matrix completion estimator (Athey et al. 2018). The primary difference between these two methods is the way they regularize the latent factor model. The results from the matrix completion estimator are similar in significance and effect size (results are available upon request).

Overall, our results are consistent across various methods and rankings, and robust to multiple falsification tests. Our estimates suggest that the rise of deep learning resulted in a divergence between *haves and have-nots*. In other words, we find that the compute divide resulted in a de-democratization of modern AI research.

**Fixed Effects Model with a Novel Compute Measure**

To establish that compute played a major role in the de-democratization, we utilize a novel measure of compute. OpenAI, a leading AI research organization, estimated compute for history's famous models and posited that from the 1960s to 2011, compute followed Moore's law. Stated differently, until 2011, available compute doubled every 2 years. However, since 2012, compute is doubling every 3.4 months. After estimating compute for prominent AI models such as AlphaGo, OpenAI extrapolated the data for each year. OpenAI's compute measure is *petaflops per second per day* or $10^{15}$ neural network operations per second per day or in total $10^{20}$ operations. OpenAI followed the KW-hr for energy tradition to create petaflop per second-day unit. This is a measure of the number of actual operations performed by the best-known models of that year. We took the log scale of the data, which ranges from $6\times10^{-8}$ petaflops per second in 2000 to $1.86\times10^{5}$ petaflops per second in 2019.[10]

We use OpenAI's annual measure of compute and interact with our *ImageNet2012* binary variable. This model aims to examine whether the modern AI, due to its compute-intensive nature, had any differential impact on different groups of organizations. To estimate the effects of compute on organizational participation in AI we use the following equation:

$$AllFirms_{it} = \alpha_i + \beta_1 X_{it} + \beta_2 * Compute_t * ImageNet2012_{it} + ImageNet2012_{it} + Compute_t + \varepsilon_{it} \quad (3)$$

Here, $\alpha_i$ is the unobserved time-invariant conference effect. $X_{it}$ is the time-variant observed factors such as the number of publications. $Compute_t$ is the OpenAI measure's annual compute measure and $ImageNet2012_{it}$ is a binary variable indicating treatment status. Our coefficient of interest is $\beta_2$. We use conference fixed effects to control for conference-level factors that could influence organizational participation. We use the same equation to estimate the effect of compute for other groups (e.g., Fortune500Tech, QS1-50) as well.

**Table 8: The effects of Compute on organizational participation in AI: FE estimates**

| | *Dependent variable:* | | | | | | |
|---|---|---|---|---|---|---|---|
| | (1) | (2) | (3) | (4) | (6) | (7) | (8) |
| | Allfirms | Fortune500Tech | QS1-50 | QS51-100 | QS101-200 | QS201-300 | QS301-500 |
| ImageNet2012 | 18.11*** | 18.72*** | 23.39*** | -0.328 | -2.48 | -4.77*** | 2.01 |
| | (2.98) | (2.24) | (3.10) | (1.45) | (1.79) | (1.47) | (1.40) |
| TotalNumofPaper | 0.239*** | 0.125*** | 0.358*** | 0.196*** | 0.196*** | 0.120*** | 0.144*** |
| | (0.005) | (0.003) | (0.005) | (0.002) | (0.003) | (0.002) | (0.002) |
| Compute | -0.0000* | 0.0000** | 0.0000 | 0.0000* | 0.0000 | 0.0000*** | 0.0000*** |
| | (0.0000) | (0.0000) | (0.0000) | (0.0000) | (0.0000) | (0.0000) | (0.0000) |
| ImageNet*Compute | 0.0007*** | 0.0007*** | 0.0001*** | -0.0003 | -0.0001*** | -0.0001*** | -0.0001*** |
| | (0.0000) | (0.0000) | (0.0000) | (0.0000) | (0.0000) | (0.0000) | (0.0000) |
| Conference FE | Yes | Yes | Yes | Yes | Yes | Yes | Yes |
| Observations | 903 | 903 | 903 | 903 | 903 | 903 | 903 |
| $R^2$ | 0.846 | 0.773 | 0.895 | 0.908 | 0.857 | 0.779 | 0.852 |

Note: Fixed effects estimates the impact of increased compute on different groups' participation in AI research after the ImageNet shock using equation 3. Standard errors are shown in parentheses and *, ** and *** denote significance at the 10%, 5%, and 1% level, respectively.

Table 8 presents the results, which are consistent with our previous models. Model 1 shows that the interaction term between Compute and ImageNet is positively significant for *AllFirms*.

[10] This is a proxy for annual *available compute*, which does not necessarily mean that all organizations need or have access to that level of computing power.

In other words, the results suggest that increased compute correlates with increased firm participation in AI since 2012. Result also suggests that one standard deviation (41700.14) increase in petaflops per second annually would increase firms' presence by 29 additional papers per year per conference. Model 2 suggests that Fortune500Tech also increased presence at the same level since the ImageNet shock. These results are consistent with our GSC method estimates.

Next, we turn our attention to different university groups. Table 8, Model 3 shows that elite universities (QS1-50) significantly increased their presence. For elite universities, one standard deviation (41700.14) increase in petaflops per second annually would increase their presence by 4 additional papers per year per conference. The treatment effect is much lower than the previous two models, which gives credence to the argument that elite universities also have limited access to compute relative to firms. Model 4 reports that universities ranked 51-100 observed significant positive impact since 2012. However, universities ranked 101-200 did not observe any noticeable impact. Moreover, the interaction term is negative for mid-tier and lower-tier universities as presented in Models 6, 7, and 8. This indicates that increased compute since 2012 negatively affected non-elite universities. Universities that are ranked 101-200, 201-300, and 301-500, each lost ground in AI research. For further robustness tests, we ran the same models with the US News 2018 ranking and produced similar results (see table D1 in Appendix D).

In sum, the results are consistent with the argument that increased compute helped large technology firms and elite universities more than mid- and lower-tier universities. Consequently, mid-tier and lower-tier universities lost research ground in top academic conferences.

**The Compute Divide between Firms and Universities: Evidence from Text data**

We now turn to text data to examine whether compute played a major role in the divide between firms and universities. First, we present the share of deep learning publications within AI conferences. To accomplish this, we took a subsample of deep learning papers from the top 10 AI conferences with an extensive list of keywords. The list of keywords was based on previous literature (Abis & Veldkamp, 2020; Alekseeva, Azar, Gine, Samila, & Taska, 2019) and an extensive consultation with deep learning researchers. The keywords list is shown in Appendix E. Figure 7 depicts that firms' share of deep learning publications has increased steadily over the years and more sharply since 2012. This increase seems to be driven particularly by Fortune500Tech firms. For most other groups, the share of deep learning papers has remained relatively stable.

**Figure 7: Share of papers within deep learning research**

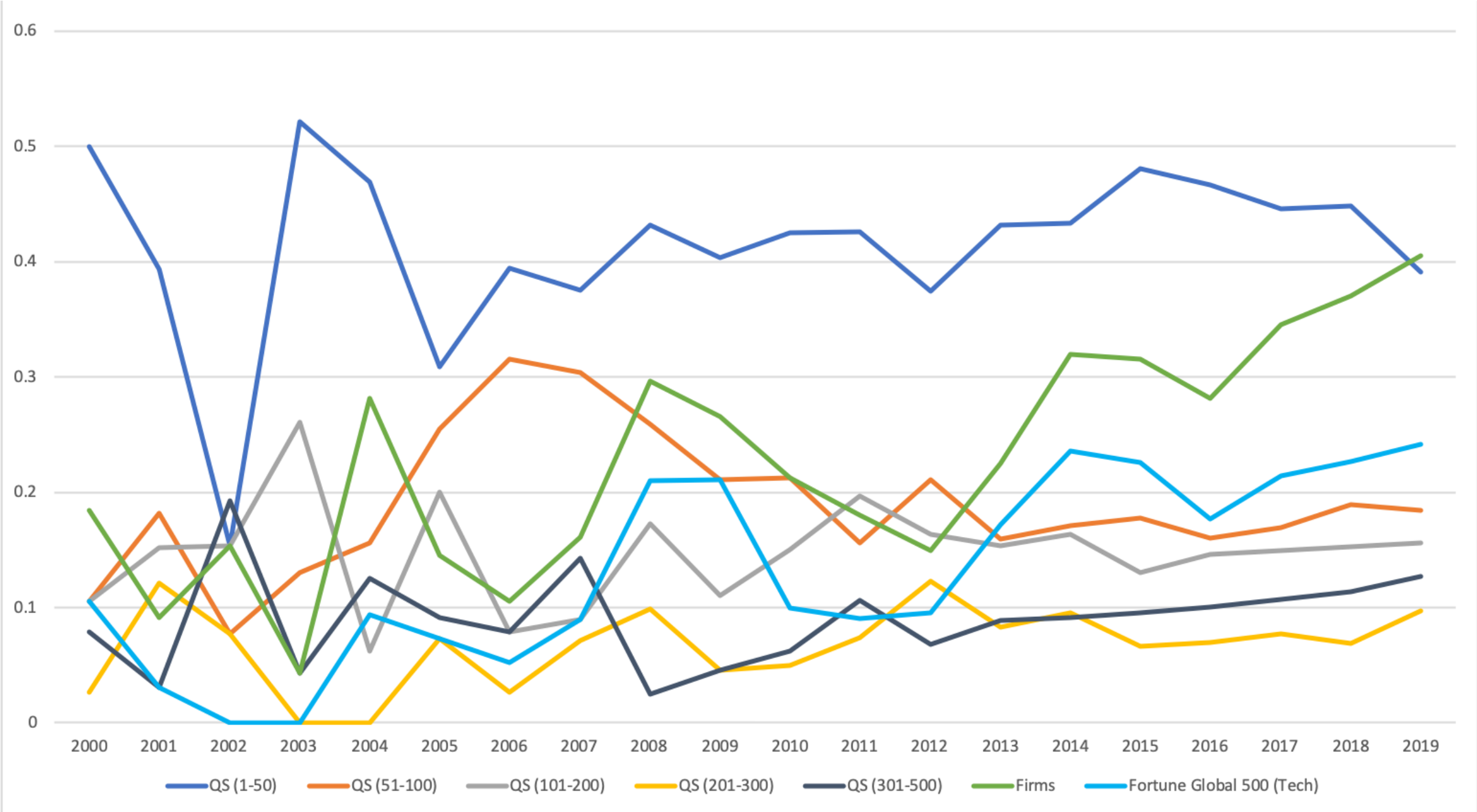

Note: This figure illustrates the share of papers that have at least one co-author from that specific group (e.g., firms, universities) within the deep learning papers.

## Machine Learning based Text Analysis: Research Focus of Different Groups

To examine how firms have been more successful in publishing AI research relative to mid-tier and lower-tier universities, we delve deeper into the text data, namely the abstracts of AI papers. We perform term frequency–inverse document frequency or TF-IDF analysis on the abstracts to gain a better understanding of different organizations' focal areas. This method quantifies the relative importance of a word to a document within a corpus. While Term Frequency (TF) considers how frequently a term appears in a document, Inverse Document Frequency (IDF) accounts for the relative importance of a term by penalizing the terms that appear in too many documents.

$$TF_w = \frac{number\ of\ occurence\ of\ term\ w}{total\ number\ of\ terms}$$

$$IDF_w = log\left(\frac{number\ of\ documents}{number\ of\ documents\ that\ have\ the\ term\ w}\right)$$

$$TFIDF_w = TF_w * IDF_w$$

To perform TF-IDF analysis, we first conduct careful preprocessing steps consisting of stemming, removal of stop words, and bi-gram formation. We then classify the AI papers of each conference into three sub-groups corresponding to the three groups of interest: Fortune500GlobalTech, QS1-50 and QS301-500. For each group, we consider the papers where at least one author was affiliated with that organization. We separately calculate TF-IDF scores for the terms for each group. Finally, we normalize the TF-IDF scores by dividing the sum of TF-IDF scores for each term by the sum of TF-IDF scores of all terms. The final score of the terms is given by the following equation:

$$S_{i,j,w} = \left(\sum_{d \in D} TFIDF_{w,d}\right) * \left(\sum_{k \in W} \sum_{d \in D} TFIDF_{k,d}\right) \qquad (4)$$

Where,

$i \in \{AAAI\}, j \in \{FortuneGlobal500\text{Tech}, QS_{1-50}, QS_{301-500}\}$, D is the set of papers, and W is the set of all terms.

Figure 8 presents TF-IDF scores for the selected keywords. The results suggest that these three groups have similarities and dissimilarities in terms of focus areas. For instance, all of these groups are mostly focused on the state of the art or SOTA research, meaning that these studies are trying to have the best result on a certain benchmark. However, these groups significantly differ in their approaches, datasets, and focus areas. For example, in deep learning, firms have a higher presence than both elite universities and non-elite universities. This is evident in keywords such as convolutional neural network, deep learning, long term short memory, and recurrent neural network. This result is consistent with recent research, which also finds that large technology firms are focusing mostly on deep learning and its commercial application (Klinger, Mateos-garcia, & Stathoulopoulos, 2020). We also find that the graphics processing unit or GPU usage is higher by large firms and top-tier universities relative to non-elite universities. Consequently, non-elite universities are more focused on traditional approaches such as support vector machine, feature selection, and Bayesian methods. The figure also suggests that non-elite universities are more concerned with computational cost and computational complexity. Taken together, this indicates that non-elite universities are still lagging behind in deep learning related research relative to large technology firms. For additional evidence, we present a similar graph for NeurIPS in Appendix C. The results from NeurIPS are qualitatively similar. Overall, our machine learning based text analysis is consistent with the argument that large firms and elite universities have an advantage in deep learning and compute-intensive research relative to non-elite universities.

**Figure 8: The difference of research focus across groups**

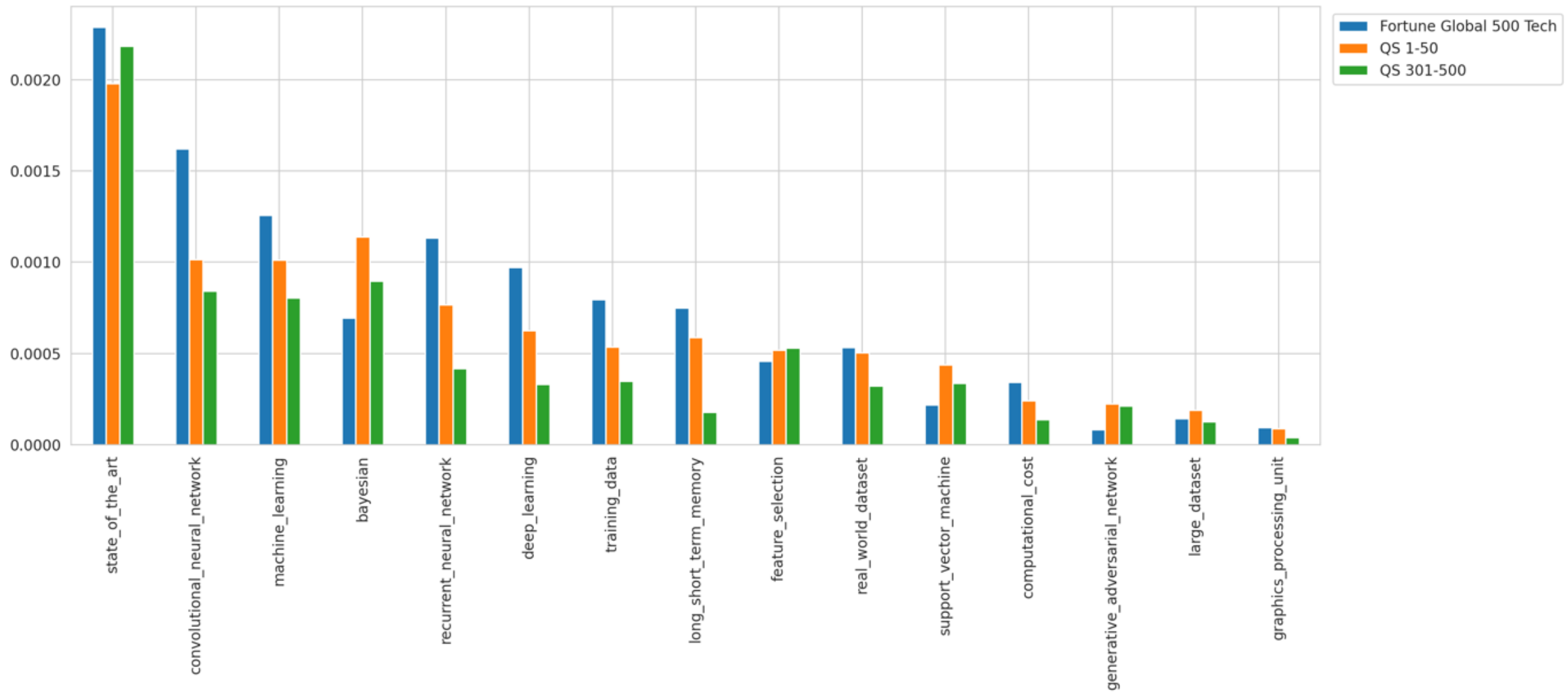

Note: This figure illustrates the normalized TF-IDF scores for three groups of organizations for the selected keywords at AAAI (2012-2018).

## Limitations

One limitation of the study is that we include only top academic venues, not all academic publications. However, the academic research agenda is often decided at such top venues (Freyne et al., 2010). Therefore, it is important to understand who is on these top venues and how that affects the future research direction. If a large number of universities lack space in these academic venues, the overall direction of AI research might be impacted in a way that is not socially optimal.

Another limitation is that our effect size for firms in the later years may be overestimated because firms' involvement in AI could affect firms' decisions to participate in other subfields of computer science. To counter this argument, we present evidence from management literature, which suggests that firms are overall *decreasing* their public research (Arora et al., 2018; Tijssen, 2004). Moreover, it is well-acknowledged that firms tend to limit publications to avoid knowledge spillover (Alexy, George, & Salter, 2013; Arora, Belenzon, & Sheer, 2017). Therefore, most firms are strategic in their publication decisions and disclose research only when that disclosure will directly or indirectly benefit them. Thus, increasing research presence in AI goes against the common trend, suggesting that firms are being strategic with their publications. While doing so, firms are inadvertently crowding out certain actors such as universities.

## Corporate Science and AI Research

Historically, a small number of large corporate labs such as AT&T, Du Pont, IBM, and Xerox played an important role in producing key innovations such as the transistor, the laser, and graphical user interface. However, over the last three decades, scholars have documented that firms have reduced presence in research (Arora et al., 2018; Larivière et al., 2018; Tijssen, 2004). Scholars argue that increasingly large firms are less involved in basic research and more interested in short-term deliverables (Tijssen, 2004). More specifically, researchers find that there is a growing division of labor between academia and industry with respect to basic science research (Arora, Belenzon, Patacconi, & Suh, 2020). For instance, Larivière et al. (2018) document that increasingly, universities have developed "near-exclusive monopoly" over publishing papers, and industry over patenting. In the same vein, Arora et al. (2020) report that firms are more reliant on universities for basic science research. However, contrary to these assertions, we find that with respect to AI, firms have increased their presence significantly. Future research should explore why this sudden increase in corporate presence is only observed in AI research.

**Implications for AI research direction**

This increased corporate presence in AI research will have a significant impact on the future direction of AI research. Innovation research suggests that industry affiliation affects the research direction of a researcher. For instance, Evans (2010) reports that industry nudges scientists towards more exploratory and speculative research. Similarly, economic theories also suggest that firms influence their employees' research directions (Aghion, Dewatripont, & Stein, 2008). Recent evidence from AI research indicates that firms and elite universities are focused more on thematically "narrow AI" (Klinger et al., 2020). Thus, increased industry activity in top academic venues might affect the overall general direction of AI research.

While industry plays an important role in academic research, it also has its limitations. Recent academic work acknowledges the importance of industry involvement, while also cautioning about the challenges of increased industry funding and participation (Irani et al., 2019). In the same vein, Gulbrandsen & Smeby (2005) find that industry funding pushes scientists towards more applied research. One major concern about industry funding and industry involvement is that it could incentivize researchers to move away from the alternative pathways that academics might find interesting. Growing concerns around increased corporate presence in AI was demonstrated in a debate titled "Academic AI Research in an Age of Industry Labs" at AAAI 2020, one of the leading AI conferences. The proposition was: *"Academic AI researchers should focus their attention on research problems that are not of immediate interest to industry."* This proposition underscores the growing tension between short-term and long-term research focus between academia and industry.

Research suggests that industry involvement that encourages increased commercialization research could negatively affect fundamental inquiry in academia (Evans, 2010a; Murray, 2010). Sociology of science research also indicates that industry sponsorship can potentially hinder the diffusion of research ideas (Evans, 2010a). One startup founder recently echoed such concerns, noting, *"[...] the tech giants are not taking a truly open source approach, and their research and engineering teams are totally disconnected. On [the] one hand, they provide black-box NLP APIs — like Amazon Comprehend or Google APIs — that are neither state-of-the-art nor flexible enough. On the other hand, they release science open source repositories that are extremely hard to use and not maintained* " (Johnson 2019). Here, the entrepreneur is conveying that large companies do not follow the open source best practices. This is even more concerning when evidence suggests that industry AI research is less reproducible compared to academic research (Gundersen & Kjensmo, 2018). In sum, mitigating bias requires transparency in the systems; however, increasingly, corporate AI research is shown to be less transparent and less reproducible. This highlights one of the potential challenges of the growing presence of industries in AI research.

On the other hand, firms' increased presence can be beneficial by increasing a division of labor between academia and industry. For instance, exploiting simultaneous discovery, Bikard, Vakili, & Teodoridis (2019) find evidence that industry affiliation increases more follow-on citation-weighted publications. The authors argue that this increased productivity is due to specialization and efficient allocation of tasks between industry and academia. This highlights how industry participation is not necessarily negative for AI's future; rather, we might observe that academia and industry will focus on different kinds of research. However, we still do not have any concrete evidence of how the industry's increased presence will affect the future of AI research. Our results open room for future discussion.

**Diversity and Innovation Outcome in AI research**

Our results are concerning given that AI technologies have shown biases against people of color due to a lack of proper training data and oversight (Buolamwini & Gebru, 2018; Koenecke et al., 2020). For instance, Bolukbasi et al. (2016) found that one popular algorithm, "word2vec", encoded existing social inequities such as gender stereotypes. Similarly, comparing major tech companies such as Amazon, Apple, Google, and Microsoft's automated speech recognition systems, Koenecke et al. (2020) found that such technologies have a higher error rate for African American speakers than for white speakers. The authors speculate that the lack of inclusive training data caused the performance gap between the two racial groups.

Overall, such cases have been found across many different technologies where either developers or datasets lack minority representation.

To limit bias and unfairness in AI research, prior research highlights the importance of having diversity within research groups (Abebe, 2018; Kuhlman et al., 2020; West et al., 2019). The core concern is that a lack of diverse perspectives leads to models and methods that do not carefully consider the consequences for minorities or vulnerable populations. Researchers argue that the lack of domain knowledge and diverse perspectives among researchers result in biased data sets, where minority presence is negligible, and the proposed models reproduce systemic bias (Kuhlman et al. 2020). This is consistent with the observation that diversity of inventions is correlated with the diversity of inventors (Koning, Samila, & Ferguson, 2019; Nielsen, Bloch, & Schiebinger, 2018). For instance, using data on biomedical patents, Koning et al. (2019) found that women-led teams are more likely to focus on female health outcomes. In AI research, it has been found that female researchers co-author studies about AI's societal consequences more than male researchers (Stathoulopoulos & Mateos-garcia, 2019). Overall, this indicates that increased diversity has the potential to reduce biases within AI research.

However, recent work in AI laments the lack of diversity in the industry. Recent empirical evidence indicates gender diversity is lower in firms than in academia. For example, Google's AI research arm has only 10% female staff, while at Facebook, the figure is 15%. Racial diversity in the industry, in particular, within large technology firms is even more concerning (Kuhlman et al., 2020). Taken together, this suggests that the AI industry is mostly homogeneous and minorities are underrepresented.

Furthermore, elite universities also have significant diversity problems. For instance, in the United States, it is well-acknowledged that elite universities are racially less diverse than mid-tier or lower-tier universities (Reardon et al., 2012). Further, elite universities in the U. S. tend to admit students from wealthy backgrounds (Chetty et al., 2019). Overall, elite universities do not represent the general population and tend to represent a privileged group of people who might not be familiar with the challenges that underprivileged groups or minorities face.

In sum, empirical evidence is consistent with the concern that both industry and elite universities have diversity problems. In light of this fact, our result poses important questions for the future of AI research. On the one hand, our results suggest that AI is increasingly shaped by large technology firms and elite universities. On the other hand, diversity is important to reduce biases in AI technologies that are being commercialized across many different domains. More research is needed to understand how the growing divergence between elite and non-elite universities could affect AI technologies.

**Policy Implications**

Our findings have several important policy implications. The results of our study highlight that resources (e.g., compute) give elite organizations an unfair advantage and create inequality and concentration of power. The prohibitive cost of compute can discourage academic researchers from pursuing certain kinds of AI research within universities and can accelerate the brain drain of academia to industry (Crumpler, 2020). Recently, Jack Clark, OpenAI's head of public policy, highlighted the importance of government intervention in measuring and understanding the societal impact of AI (Clark, 2019). He argued that the U.S. government should step up to help certain universities with resources. Similarly, a group of Stanford computer scientists also argued for the "National Research Cloud," which will ensure affordable access to compute for academics (Walsh, 2020). Our results find the first concrete evidence that government intervention may be necessary to reduce the "compute divide."

Our results also suggest that data is an important input in AI knowledge production. Prior research demonstrates that access to data democratizes science (Nagaraj et al., 2020). Shared public datasets that can help to train and test AI models will be particularly beneficial for resource-constrained organizations. We posit that by releasing publicly owned data, governments can help non-elite universities and startups in the AI research race.

The increased concentration of power in AI research has important implications for regulators as well. For instance, if AI research requires significant upfront costs in terms of hiring human capital, acquiring expensive datasets, and compute, that could increase the entry barriers for startups. Thus, large companies will be insulated from potential disruptions from new startups. This could also lead to a concentration of power at the hands of a few actors at the industry level.

**Conclusion**

AI is one of the most consequential technologies of our time, and it is well-acknowledged that democratizing AI will benefit a large number of people. Exploiting the sudden rise of deep learning due to an unanticipated usage of GPUs since 2012, we find that AI is increasingly being shaped by a few actors, and these actors are mostly affiliated with either large technology firms or elite universities. To do so, we use 171,394 peer-reviewed papers from 57 major computer science conferences. Our data also shows that there is a marked difference in the quantity of production of AI knowledge between elite and non-elite universities. Consequently, we find that hundreds of mid-tier and lower-tier universities are being crowded out of the AI research space. These findings are consistent with the emphasis that access to compute is playing a major role in this divergence. Additionally, we document that historically Black and Hispanic serving institutions have very limited presence in top AI conferences.

Further, using machine learning based text analysis, we provide evidence that the divergence between firms and universities is occurring partly due to uneven access to computing power. We call this unequal access to compute between firms and certain universities "compute divide.” This has important implications for public policy and technology governance since AI affects our shared future significantly. To truly “democratize” AI, a concerted effort by policymakers, academic institutions, and firm-level actors is needed to tackle the compute divide.

We contribute to the growing literature on the role of specialized equipment and materials on knowledge production (Ding et al., 2010; Stephan, 2012; Teodoridis, 2018) by demonstrating that lack of access to certain resources can de-democratize a scientific field. We also contribute to the innovation literature on corporate science by documenting that contrary to recent evidence (Arora et al., 2018; Larivière et al., 2018), firms are increasing research presence in AI.

## Appendix A

**Table A1: List of Conferences (Bold indicates treated conferences):**

| Area | Abbreviation | Name |
|---|---|---|
| **Artificial Intelligence** | **AAAI** | **Association for the Advancement of Artificial Intelligence** |
| **Artificial Intelligence** | **IJCAI** | **International Joint Conferences on Artificial Intelligence** |
| **Computer Vision** | **CVPR** | **Conference on Computer Vision and Pattern Recognition** |
| **Computer Vision** | **ECCV** | **European Conference on Computer Vision** |
| **Computer Vision** | **ICCV** | **International Conference on Computer Vision** |
| **Machine Learning & Data Mining** | **ICML** | **International Conference on Machine Learning** |
| **Machine Learning & Data Mining** | **KDD** | **Conference on Knowledge Discovery and Data Mining** |
| **Machine Learning & Data Mining** | **NeurIPS** | **Conference on Neural Information Processing Systems** |
| **NLP** | **ACL** | **Association for Computational Linguistics** |
| **NLP** | **EMNLP** | **Empirical Methods in Natural Language Processing** |
| The Web & Information Retrieval | WWW | International World Wide Web Conference |
| The Web & Information Retrieval | SIGIR | Special Interest Group on Information Retrieval |
| Computer architecture | ASPLOS | International Conference on Architectural Support for Programming Languages and Operating Systems |
| Computer architecture | ISCA | International Symposium on Computer Architecture |
| Computer architecture | MICRO | International Symposium on Microarchitecture |
| Computer Networks | SIGCOMM | Association for Computing Machinery's Special Interest Group on Data Communications |
| Databases | SIGMOD | Special Interest Group on Management of Data |
| Databases | VLDB | International Conference on Very Large Data Bases |
| Design automation | DAC | Design Automation Conference |
| Design automation | ICCAD | International Conference on Computer Aided Design |
| Embedded & Real-Time Systems | EMSOFT | International Conference on Embedded Software |
| Embedded & Real-Time Systems | RTAS | IEEE Real-Time and Embedded Technology and Applications Symposium |
| Embedded & Real-Time Systems | RTSS | IEEE Real-Time Systems Symposium |
| High-Performance Computing | HPDC | High-Performance Parallel and Distributed Computing |
| High-Performance Computing | ICS | International Conference on Supercomputing |
| High-Performance Computing | SC | International Conference for High Performance Computing, Networking, Storage and Analysis |
| Measurement & Perf. Analysis | IMC | Internet Measurement Conference |
| Measurement & Perf. Analysis | SIGMETRICS | Special Interest Group on Measurement and Evaluation |
| Mobile Computing | MobiCom | Annual International Conference on Mobile Computing and Networking |
| Mobile Computing | MobiSys | International Conference on Mobile Systems, Applications, and Services |
| Mobile Computing | SenSys | Conference on Embedded Networked Sensor Systems |
| Operating Systems | OSDI | Operating Systems Design and Implementation |
| Operating Systems | SOSP | Symposium on Operating Systems Principles |
| Operating Systems | EuroSys | European Conference on Computer Systems |
| Programming Languages | PLDI | Programming Language Design and Implementation |
| Programming Languages | POPL | Symposium on Principles of Programming Languages |
| Security | ACMCCS | ACM Conference on Computer and Communications Security |
| Software Engineering | ICSE | International Conference on Software Engineering |
| Software Engineering | SIGSOFT/FSE | Symposium on the Foundations of Software Engineering |
| Algorithms & Complexity | FOCS | Symposium on Foundations of Computer Science |
| Algorithms & Complexity | SODA | Symposium on Discrete Algorithms |
| Algorithms & Complexity | STOC | Symposium on Theory of Computing |
| Cryptography | EUROCRYPT | Annual International Conference on the Theory and Applications of Cryptographic Techniques |
| Logic & Verification | CAV | International Conference on Computer-Aided Verification |
| Logic & Verification | LICS | Symposium on Logic in Computer Science |
| Comp. Bio & Bioinformatics | RECOMB | International Conference on Research in Computational Molecular Biology |
| Computer Graphics | TOG | ACM Transactions on Graphics |
| Human-Computer Interaction | CHI | Conference on Human Factors in Computing Systems |
| Human-Computer Interaction | UBICOMP | International Joint Conference on Pervasive and Ubiquitous Computing |
| Human-Computer Interaction | UIST | Symposium on User Interface Software and Technology |
| Robotics | ICRA | International Conference on Robotics and Automation |
| Robotics | IROS | International Conference on Intelligent Robots and Systems |
| Robotics | RSS | Robotics: Science and Systems |
| Visualization | IEEEVIS | IEEE Transactions on Visualization and Computer Graphics |
| Visualization | IEEEVR | The IEEE Conference on Virtual Reality and 3D User Interface |
| Visualization | INFOCOM | Conference on Computer Communications |

**Table A2:**
**List of 46 firms as mentioned in Fortune500 Global Technology firms**

| |
|---|
| Apple |
| Samsung Electronics |
| Amazon.com |
| Hon Hai Precision Industry |
| Alphabet |
| Microsoft |
| Huawei Investment & Holding |
| Hitachi |
| IBM |
| Dell Technologies |
| Sony |
| Panasonic |
| Intel |
| LG Electronics |
| JD.com |
| HP |
| Cisco Systems |
| Lenovo Group |
| Facebook |
| Honeywell International |
| Mitsubishi Electric |
| Pegatron |
| Alibaba Group Holding |
| Oracle |
| Fujitsu |
| Accenture |
| Canon |
| Midea Group |
| Toshiba |
| Tencent Holdings |
| Quanta Computer |
| Taiwan Semiconductor Manufacturing |
| China Electronics |
| Compal Electronics |
| Hewlett Packard Enterprise<br>Schneider Electric<br>Wistron<br>SK Hynix<br>SAP<br>Onex<br>Nokia<br>NEC<br>Flex<br>LG Display<br>Qingdao Haier<br>Ericsson |

## Appendix B

### Figure B1: Panel data structure of all the conferences

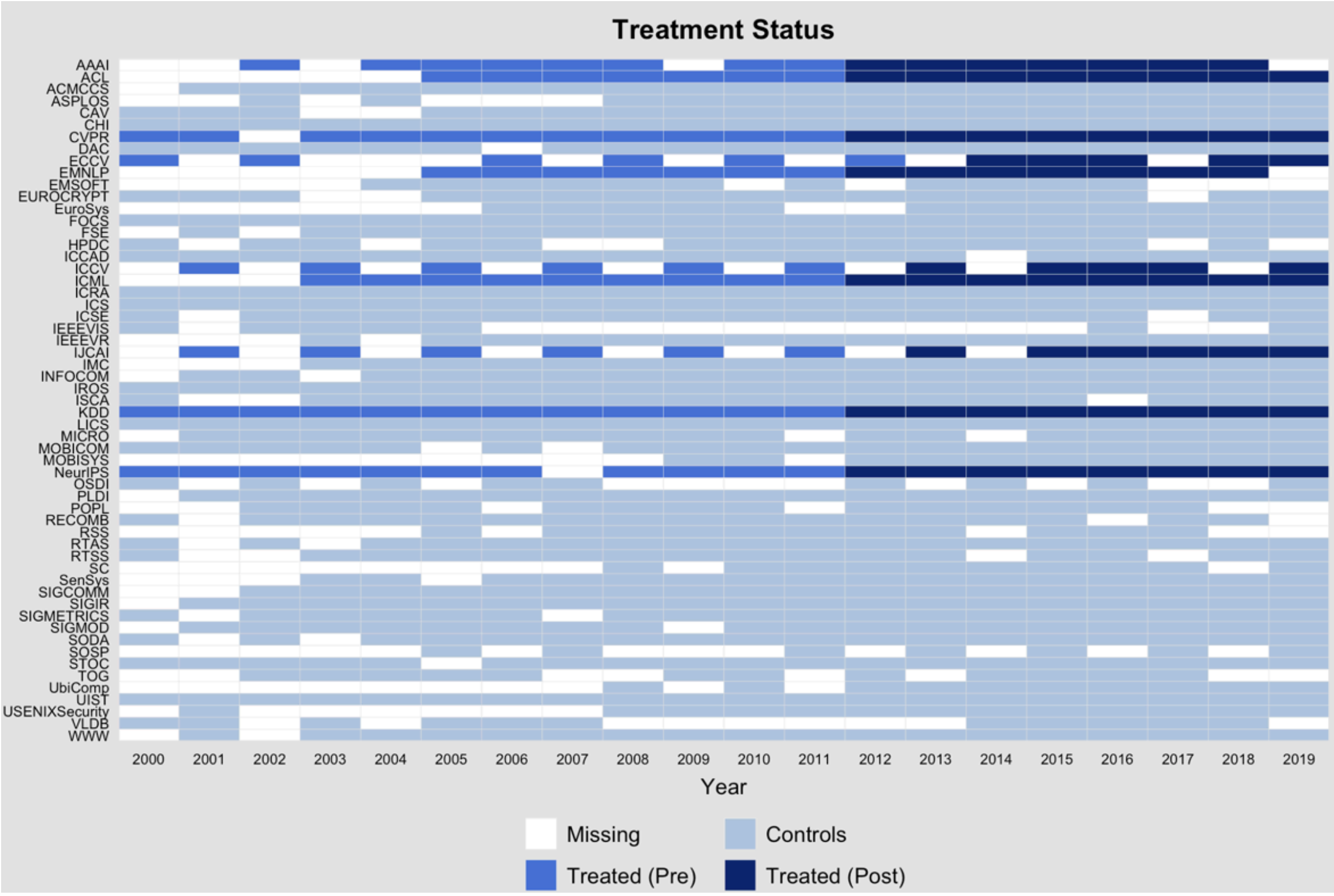

Note: this figure depicts the structure of the panel data of the 57 conferences. Each box represents one observation in the data. A few conferences were biannual such as ICCV, ECCV, IJCAI.

**Figure B2: *AllFirms* for each conference over time**

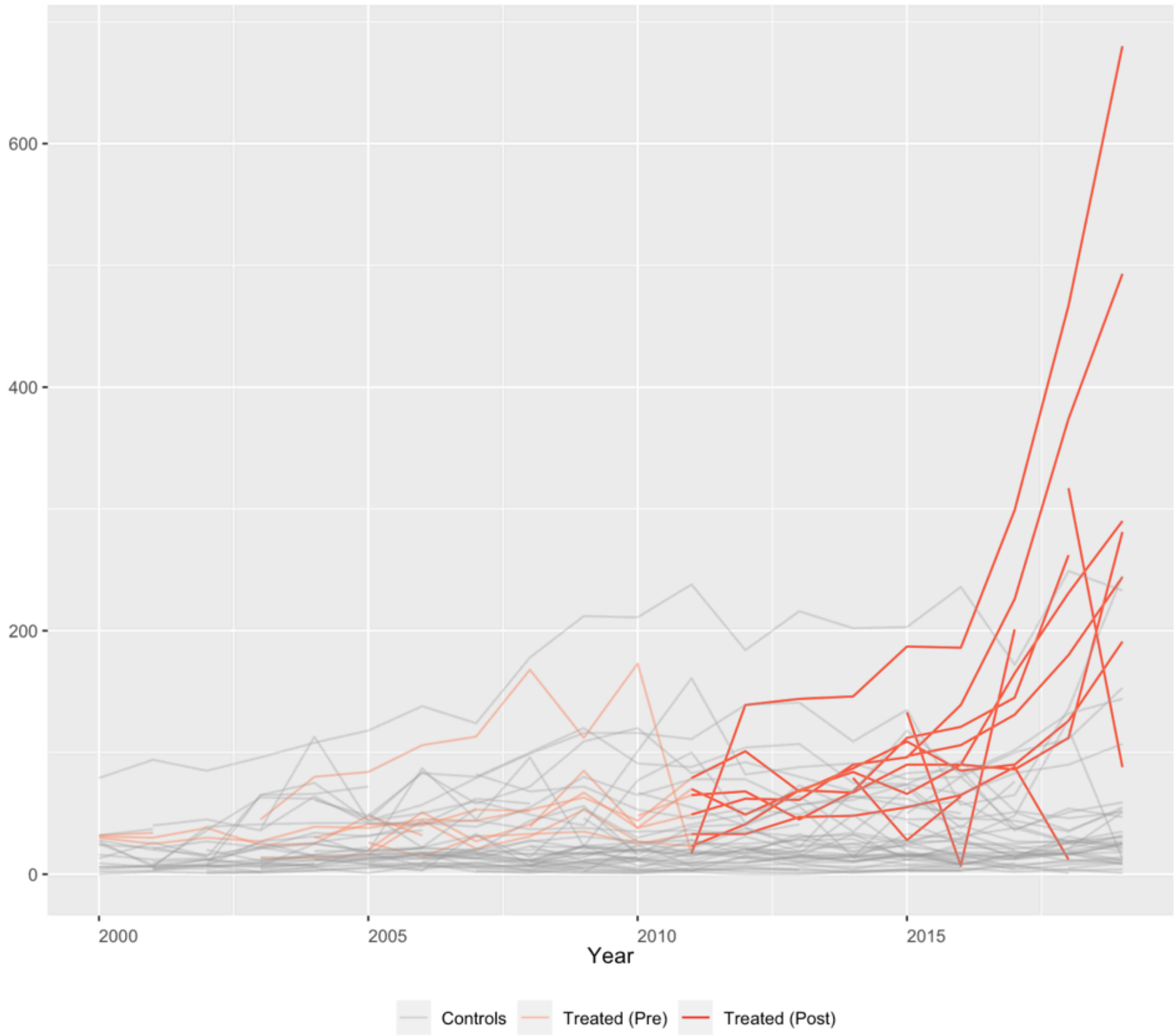

Note: This shows the *AllFirms* variable for both AI and non-AI conferences over time. AI conferences seem to diverge from the rest of the crowd only after 2012.

## <u>Appendix C:</u>

**Figure C1: The difference of research focus across groups**

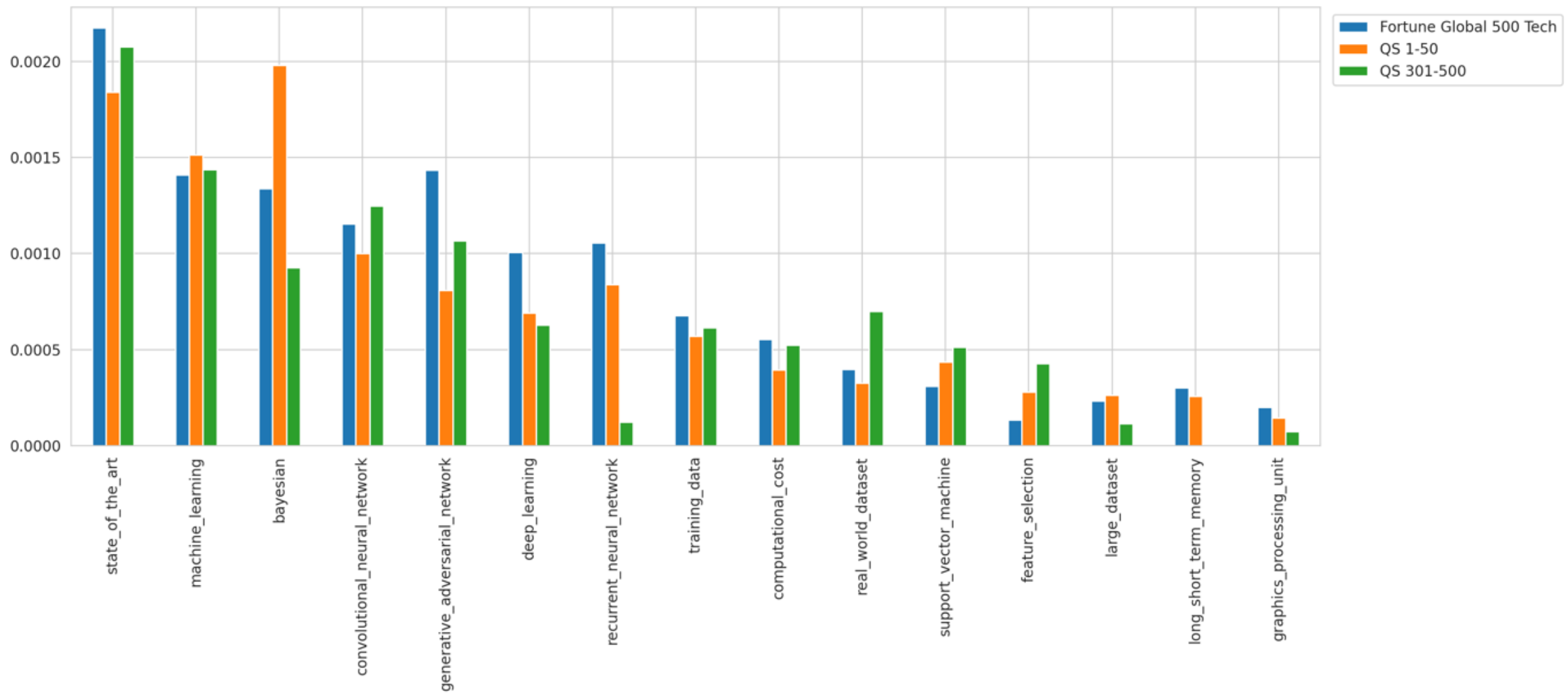

Note: This figure illustrates the normalized TF-IDF scores for three groups of organizations for the selected keywords at NeurIPS (2012-2018).

## Appendix D:

## Table D1: The effects of Compute on organizational participation in AI: Fixed effects estimates (US News 2018 ranking)

| | *Dependent variable:* | | | | | |
|---|---|---|---|---|---|---|
| | (1) | (2) | (3) | (4) | (6) | (7) |
| | Allfirms | Fortune500Tech | US1-50 | US51-100 | US201-300 | US301-500 |
| ImageNet2012 | 18.11*** | 18.72*** | 8.42*** | 6.83*** | -5.27*** | 3.39** |
| | (2.98) | (2.24) | (3.21) | (1.49) | (1.35) | (1.60) |
| TotalNumofPaper | 0.239*** | 0.125*** | 0.317*** | 0.207*** | 0.116*** | 0.165*** |
| | (0.005) | (0.003) | (0.005) | (0.003) | (0.002) | (0.002) |
| Compute | -0.0000* | 0.0000** | 0.0000 | 0.0000 | 0.0000*** | 0.0000*** |
| | (0.0000) | (0.0000) | (0.0000) | (0.0000) | (0.0000) | (0.0000) |
| ImageNet*Compute | 0.0007*** | 0.0007*** | 0.0002*** | -0.0003*** | -0.0001*** | -0.0001*** |
| | (0.0000) | (0.0000) | (0.0000) | (0.0000) | (0.0000) | (0.0000) |
| Conference FE | Yes | Yes | Yes | Yes | Yes | Yes |
| Observations | 903 | 903 | 903 | 903 | 903 | 903 |
| $R^2$ | 0.846 | 0.773 | 0.849 | 0.915 | 0.779 | 0.851 |

Note: The table tests the impact of increased compute on different groups' participation in AI research after the ImageNet shock using equation 3. Standard errors are shown in parentheses and *, **, and *** denote significance at the 10%, 5%, and 1% level, respectively.

## Table D2: The Collaboration Strategies of firms: GSC method estimates (US News 2018 ranking)

| | *Dependent variable:* | | | | | | |
|---|---|---|---|---|---|---|---|
| | (1) | (2) | (3) | (4) | (5) | (6) | (7) |
| | AllFirms | Fortune500Tech | QS1-50 | QS51-100 | QS101-200 | QS201-300 | QS301-500 |
| ImageNet2012 | 25.72 | 22.3*** | 19.14*** | -5.96*** | -5.78* | -11.95*** | -7.126*** |
| | (19.63) | (1.329) | (3.42) | (1.391) | (3.557) | (1.155) | (0.80) |
| TotalNumOfPaper | 0.083*** | 0.035*** | 0.185*** | 0.114*** | 0.116*** | 0.089*** | 0.094*** |
| | (0.004) | (0.003) | (0.005) | (0.003) | (0.006) | (0.002) | (0.002) |
| Conference FE | Yes | Yes | Yes | Yes | Yes | Yes | Yes |
| Year FE | Yes | Yes | Yes | Yes | Yes | Yes | Yes |
| Observations | 903 | 903 | 903 | 903 | 903 | 903 | 903 |
| Treated Conference | 10 | 10 | 10 | 10 | 10 | 10 | 10 |
| Control Conference | 47 | 47 | 47 | 47 | 47 | 47 | 47 |

Note: This table estimates the 2012's ImageNet shock on firms' collaboration strategies in AI conferences. We used 10 AI conferences as the treatment group and 47 non-AI conferences as the control group. Standard errors and 95% confidence intervals are computed using 2,000 parametric bootstraps blocked at the conference level. Standard errors are shown in parentheses and *, **, and *** denote significance at the 10%, 5%, and 1% level, respectively.

**Appendix E:**

List of Keywords used to select the subset of papers related to deep learning:

Attention Mechanism, Auto Encoder, Auto Regressive Model, Autoencoder, BERT, Back Propagation, Backpropagation, Bidirectional Encoder Representations, Boltzmann Machine, CNN, CNTK, CUDA, Caffe, Caffe2, Chainer, Convolutional Network, DL4J, DLib, DNN, Deep Architecture, Deep Autoencoder, Deep Belief Network, Deep Convolutional, Deep Deterministic Policy Gradient, Deep Embedding, Deep Encoder Decoder, Deep Generative Model, Deep Generative Network, Deep Hashing Method, Deep Learning, Deep Linear Network, Deep Metric Learning, Deep Model, Deep Network, Deep Probabilistic Model, Deep Q Learning, Deep Q Network, Deep Recurrent Network, Deep Reinforcement Learning, Deep Representation Learning, Deep Supervised Hashing, Deeplearning4j, Depth Wise Convolution, DyNet, Encoder Decoder, GAN, GPU, GRU, Gated Recurrent Unit, Generative Adversarial Net, Generative Adversarial Network, GloVe, Gluon, Gradient Descent, Graphics Processing Unit, GraspNET, Hopfield Network, Keras, LSTM, Lasagne, Liquid State Machine, Long Short Term Memory, Max Pooling, Microsoft Cognitive Toolkit, Multilayer Perceptron, Mxnet, NVIDIA, Neural Architecture, Neural Language Model, Neural Machine Translation, Neural Model, Neural Net, Neural Network, Neural Style Transfer, Neural Turing Machine, ONNX, OpenNN, Opencl, Opencv, PaddlePaddle, Pytorch, RNN, Radial Basis Function Network, ReLU, Recurrent Network, Resnet, Seq2seq, Sonnet, Spiking Neural Network, TPU, Tensor Processing Unit, Tensorflow, Tflearn, Theano, Titan X, Torch, Transfer Learning, VAE, Variational Autoencoder, Word2vec, cuDNN

**The rise of deep learning and compute in AI research since 2012:**

In this section, we document the rise of compute-intensive research in AI. We calculate the normalized TF-IDF scores using equation 4. Figure E1 presents the results for NeurIPS. The result suggests that deep learning related keywords have seen significant usage since 2012. For instance, convolutional neural network, recurrent neural network, long short-term memory barely used before 2012. Since 2012, these keywords have been used extensively. Similarly, the keyword generative adversarial network has not been used before 2012 but received prominence afterward. Traditional approaches such as Bayesian methods, support vector machine, and feature selection are becoming less popular since 2012.

**Figure E1: The rise of compute-intensive research in NeurIPS**

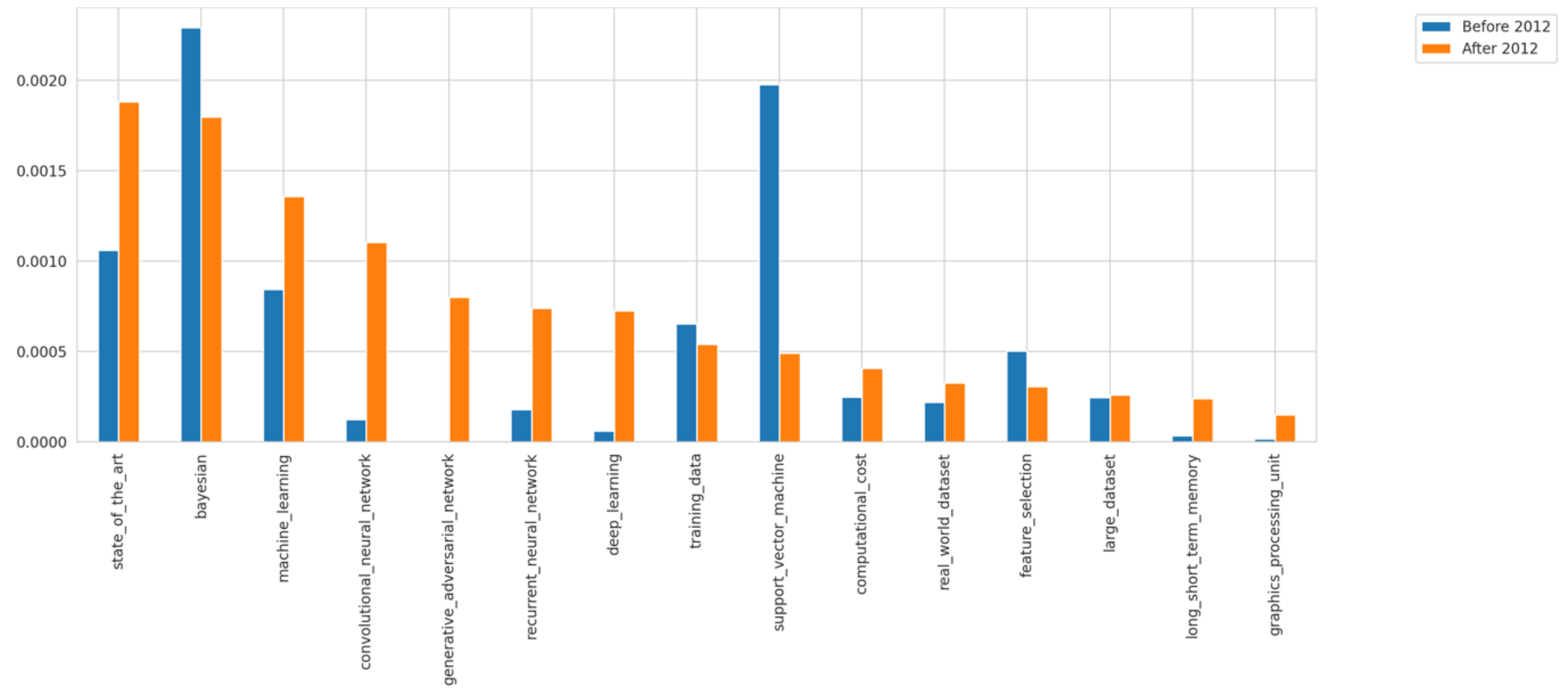

Note: This figure illustrates the normalized TF-IDF scores for before and after 2012 for selected keywords at NeurIPS (2000-2018).

The pattern is similar for AAAI which is presented in Figure E2. We observe that deep learning is more widely used and traditional methods are getting less popular.

**Figure E2: The rise of compute-intensive research in AAAI**

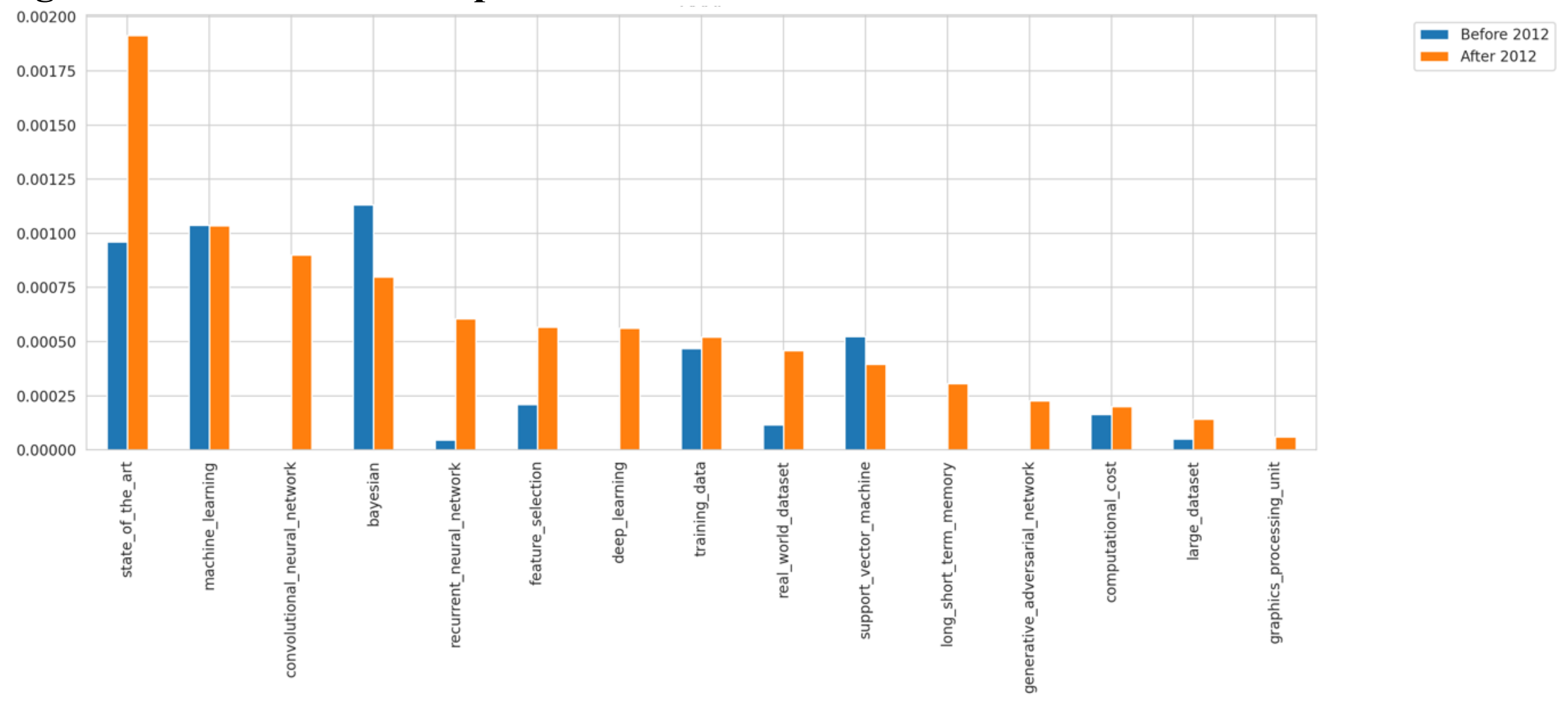

Note: This figure illustrates the normalized TF-IDF scores for before and after 2012 for selected keywords at AAAI (2000-2018).

Taken together, these two figures suggest that since 2012, AI research has become more deep learning dependent, and thus, compute-intensive.